\documentclass[12pt,preprint]{aastex}
\newcommand{\uJy}{\ensuremath{\mu{\rm Jy}}}
\newcommand{\um}{\ensuremath{\mu{\rm m}}}

\shorttitle{ATLAS Radio Observations of ELAIS-S1}
\shortauthors{Middelberg et al.}

\begin{document}


\title{Deep ATLAS Radio Observations of the ELAIS-S1/{\it Spitzer}
Wide-Area Infrared Extragalctic field}


\author{Enno Middelberg}
\affil{Astronomischies Institut der Universit\"at Bochum, Universit\"atsstr. 150, 44801 Bochum, Germany; middelberg@astro.rub.de}

\author{Ray P. Norris}
\affil{Australia Telescope National Facility, PO Box 76, Epping NSW 1710, Australia}

\author{Tim J. Cornwell}
\affil{Australia Telescope National Facility, PO Box 76, Epping NSW 1710, Australia}

\author{Maxim A. Voronkov}
\affil{Australia Telescope National Facility, PO Box 76, Epping NSW 1710, Australia}

\author{Brian D. Siana}
\affil{Spitzer Science Center, California Institute of Technology, MS 220-6, Pasadena, CA 91125}

\author{Brian J. Boyle}
\affil{Australia Telescope National Facility, PO Box 76, Epping NSW 1710, Australia}

\author{Paolo Ciliegi}
\affil{INAF, Osservatorio Astronomico di Bologna, Via Ranzani 1, I-40127 Bologna, Italy}

\author{Carole A. Jackson}
\affil{Australia Telescope National Facility, PO Box 76, Epping NSW 1710, Australia}

\author{Minh T. Huynh}
\affil{Spitzer Science Center, California Institute of Technology, MS 220-6, Pasadena, CA 91125}

\author{Stefano Berta}
\affil{Dipartimento di Astronomia, Universit\`a di Padova, Vicolo dell'Osservatorio 2, 35122 Padova, Italy}

\author{Stefano Rubele}
\affil{Dipartimento di Astronomia, Universit\`a di Padova, Vicolo dell'Osservatorio 2, 35122 Padova, Italy}

\author{Carol J. Lonsdale}
\affil{Center for Astrophysics and Space Sciences, University of California at San Diego, 9500 Gilman Drive, La Jolla, CA 92093-0424}

\author{Rob J. Ivison}
\affil{UK Astronomy Technology Centre, Royal Observatory, Blackford Hill, Edinburgh EH9 3HJ}

\author{Ian Smail}
\affil{Institute for Computational Cosmology, Durham University, South Road, Durham DH1 3LE, UK}

\author{Seb J. Oliver}
\affil{Astronomy Centre, CPES, University of Sussex, Falmer, Brighton BN1 9QJ, UK}

%\author{C. D. Biemesderfer\altaffilmark{4,5}}
%\affil{National Optical Astronomy Observatories, Tucson, AZ 85719}
%\email{aastex-help@aas.org}
%
%\and
%
%\author{R. J. Hanisch\altaffilmark{5}}
%\affil{Space Telescope Science Institute, Baltimore, MD 21218}

%\altaffiltext{1}{Visiting Astronomer, Cerro Tololo Inter-American Observatory.
%CTIO is operated by AURA, Inc.\ under contract to the National Science
%\altaffiltext{2}{Society of Fellows, Harvard University.}
%\altaffiltext{3}{present address: Center for Astrophysics,
%    60 Garden Street, Cambridge, MA 02138}
%\altaffiltext{4}{Visiting Programmer, Space Telescope Science Institute}
%\altaffiltext{5}{Patron, Alonso's Bar and Grill}


\begin{abstract}

We have conducted sensitive ($1\,\sigma<30\,\uJy$) 1.4\,GHz radio
observations with the Australia Telescope Compact Array of a field
largely coincident with infrared observations of the {\it Spitzer}
Wide-Area Extragalactic Survey. The field is centred on the European
Large Area ISO Survey S1 region and has a total area of
$3.9^\circ$. We describe the observations and calibration, source
extraction, and cross-matching to infrared sources. Two catalogues are
presented; one of the radio components found in the image and one of
radio sources with counterparts in the infrared and extracted from the
literature. 1366 radio components were grouped into 1276 sources, 1183
of which were matched to infrared sources. We discover 31 radio
sources with no infrared counterpart at all, adding to the class of
Infrared-Faint Radio Sources.

\end{abstract}



\section{Introduction}

In this paper describing early results from the Australia Telescope
Large Area Survey (ATLAS), we present a 1.4\,GHz survey of the
European Large Area ISO Survey (ELAIS) S1 field
(\citealt{Oliver2000}). This is the second survey paper describing
results from ATLAS, and is complementary to the paper by
\cite{Norris2006a}, which describes observations of the Chandra Deep
Field South.

ATLAS is an ambitious study of galaxies and their evolution since
z$\gtrsim$3, using predominantly observations in the radio regime. Two
large areas (about 3.5\,deg$^2$ each) have been surveyed with high
sensitivity, using the Australia Telescope Compact Array (ATCA) at
1.4\,GHz, to complement multi-wavelength observations in the infrared
with the {\it Spitzer} Space Telescope. The immediate goals of the
observations are, in brief, to determine whether the radio-infrared
relation holds at high redshifts; to search for overdensities of
high-z ULIRGs which mark the positions of protoclusters in the early
universe; to trace the radio luminosity function to high redshifts; to
determine the relative contribution of starbursts and AGN to the
overall energy density of the universe; and to open a new parameter
space to allow for serendipitous discovery of rare sources. However,
surveys such as this have proven in the past to have a substantial
impact on longer time-scales, when they are used as a resource in a
broad variety of studies.

It was decided early on in this project to observe two separate sky
regions, rather than one larger area, to exclude cosmic variance which
might affect the results. Both fields extend beyond $2^\circ$ in one
dimension, which is sufficient to sample structures at any one
redshift which have evolved to more than 150\,Mpc at the present
epoch. Nevertheless, such surveys are still prone to cosmic
variance. The CDFS is known to contain some large-scale structures
(\citealt{Vanzella2005} and references therein), predominantly at
redshifts of 0.73 and 1.1. These structures are not obvious clusters,
but ``sheets'' in the original CDFS. This finding demonstrates the
need to sample large areas for an unbiased view of galaxy formation.

It should be noted that the ELAIS-S1 field has been also observed by
\cite{Gruppioni1999} using the ATCA in June 1997, reaching a fairly
uniform rms of 80\,\uJy\ across the observed area, a factor of three
higher than the average rms in our current observations. Also, the
area they observed is slightly larger than the field described
here. We have cross-matched their sources to ours and briefly discuss
the results.

This paper is organised as follows. Section~\ref{sec:obs} describes
the observations and Section~\ref{sec:images} details the source
extraction process and the cross-identification of radio sources with
the SWIRE catalogue. Section~\ref{sec:cats} provides a description of
the catalogues and the literature search for counterparts in other
surveys. Section~\ref{sec:class} gives a description of the
classification and a few individual sources are described in more
detail in Section~\ref{sec:individual}. We provide a short analysis of
the radio-infrared relation in Section~\ref{sec:r-ir} and our
conclusions in Section~\ref{sec:conc}.

\section{Observations}
\label{sec:obs}

\begin{figure}
\includegraphics[width=\linewidth]{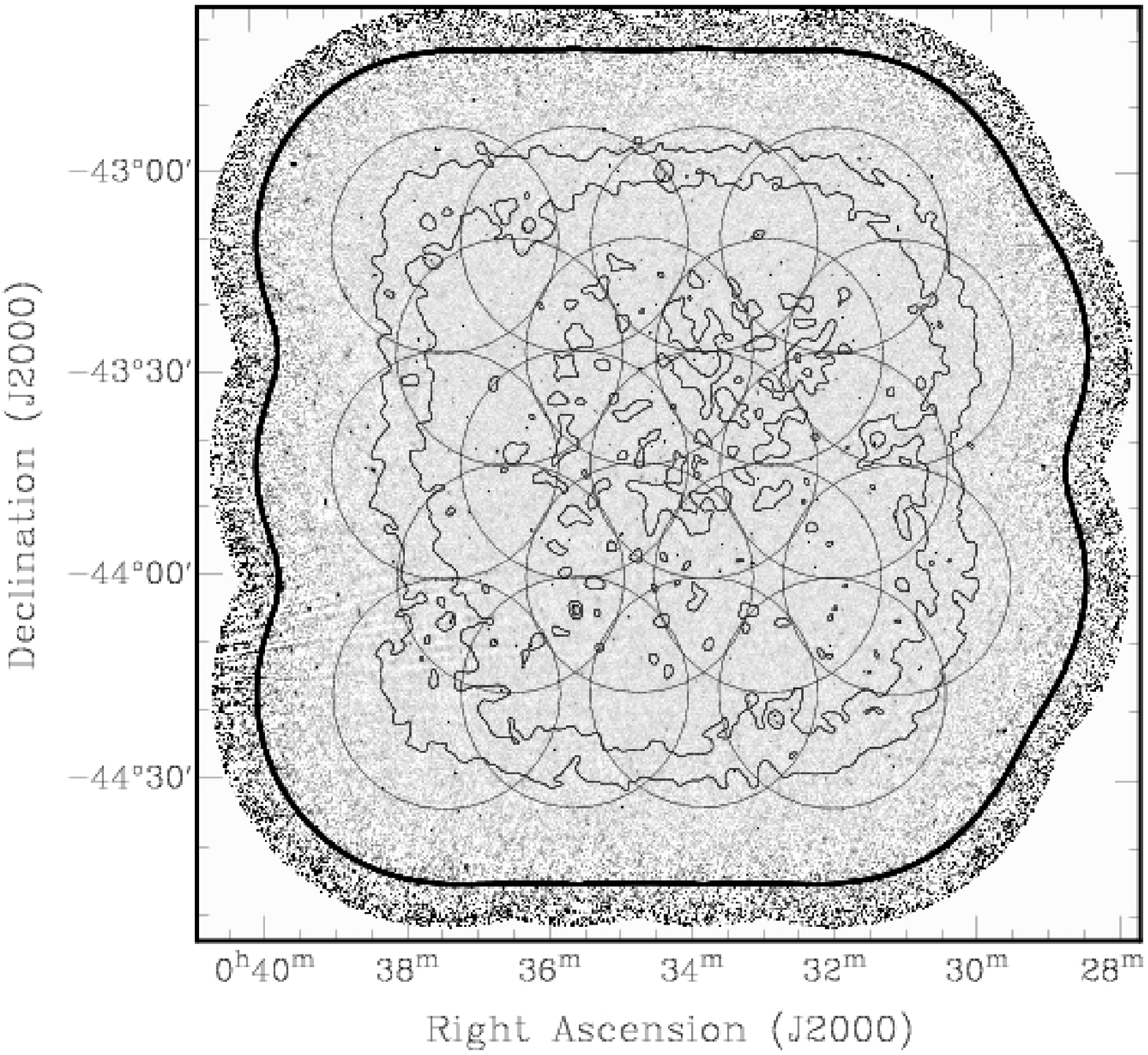}
\caption{An overview of the observed area. The circles indicate the
20 antenna pointings and the FWHM of the primary beams. The thin
contours show noise levels of 25\,\uJy, 35\,\uJy, and 45\,\uJy, as
calculated by SExtractor (\citealt{Bertin1996}). The thick contour
indicates where the predicted sensitivity is 250\,\uJy\ and marks the
area which we have analysed. The image has been clipped where the
response of the antenna primary beams has dropped to below 3\,\% of
its peak value.}
\label{fig:overview}
\end{figure}

As of December 2006 we have completed about 50\,\% of the planned
observations of the ELAIS-S1 field. The sensitivity here is slightly
higher than in the CDFS, partly because of slightly longer integration
times (between 10.5\,h per pointing and 13.5\,h per pointing, compared
to 8.2\,h per pointing over most of the CDFS), and partly because
there is a strong interfering source in the CDFS field. However, a
3.8\,Jy source (PKS\,0033-44) limits the dynamic range of the ELAIS-S1
observations even though it is well outside our pointings. An overview
of the observed area is reproduced in Figure~\ref{fig:overview}.

\subsection{Radio observations}

The radio observations were carried out on 27 separate days between 9
January 2004 and 24 June 2005 with the Australia Telescope Compact
Array (ATCA), with a total net integration time on the pointings of
231\,h, in a variety of configurations to maximise the $(u,v)$
coverage (Table~\ref{tab:obs}). However, $(u,v)$ coverage is probably
not a crucial factor in aperture synthesis when the field is dominated
by point sources as in our case. 3.89\,deg$^2$ in the ELAIS-S1 field
were analysed (this is the total area in the mosaic where the primary
beam response is $>10$\,\%) in a mosaic consisting of 20 overlapping
pointings (Table~\ref{tab:coords}). The full width at half maximum
(FWHM) of the primary beams at 1.4\,GHz is $35^\prime$. The pointings
were observed for one minute each, and the calibrator 0022-423 was
observed after each cycle of 20 pointings for two minutes. Amplitude
calibration was done using PKS\,1934-638 as a primary calibrator,
which was observed for 10\,min before or after each observing run. It
was assumed to have a flux density of 15.012\,Jy at 1.34\,GHz and
14.838\,Jy at 1.43\,GHz, corresponding to the centres of the two ATCA
frequency bands. Each band had a bandwidth of 128\,MHz over 33
channels, so the total observing bandwidth was 256\,MHz. In the
observation in early 2004, the higher band was only slightly affected
by terrestrial radio-frequency interference (RFI), but this
deteriorated in 2004, requiring considerable effort to edit the data
properly to avoid losing a large fraction of good data. The lower band
was mostly free of RFI and required little editing. In the early
stages of the project in 2004, only pointings 1-12 were observed (the
upper three rows of circles in Figure~\ref{fig:overview}), but the
surveyed area was extended in 2005 by adding pointings 13-20 to the
field. The new pointings were initially observed for a longer time to
catch up with the older pointings, resulting in a different $(u,v)$
coverage and a little more integration time. Pointings 1-12 have net
integration times of 10.5\,h per pointing, whereas pointings 13-20
have net integration times of 13.5\,h per pointing. After editing, the
predicted noise level is 22\,\uJy\ in the centre of the
mosaic. Towards the image edges, the noise level increases due to
primary beam attenuation.

\subsubsection{Calibration}

The data were calibrated using Miriad (\citealt{Sault1995}) standard
procedures, following recommendations for high dynamic range
imaging. The raw data come in RPFITS format, and are converted into
the native Miriad format using ATLOD. ATLOD discarded every other
frequency channel (which are not independent from one another, hence
no information is lost) and flagged one channel in the
higher-frequency band which contained a multiple of 128\,MHz, and thus
was affected by self-interference at the ATCA. We also did not use the
channels at either end of the band where the sensitivity dropped
significantly. The resulting data set contained two frequency bands,
with 13 channels and 12 channels respectively, all of which are 8\,MHz
wide, and so the total net bandwidth in the data was
25$\times$8\,MHz=200\,MHz.

The data were bandpass-calibrated to prepare for RFI removal with
Pieflag (\citealt{Middelberg2006a}). Pieflag derives baseline-based
statistics from a channel which is free of, or only very slightly
affected by, RFI, and searches the other channels for outliers and
sections of high noise. It is therefore important to
bandpass-calibrate the data before using it. Pieflag eliminated all
RFI-affected data which would have been flagged in a visual
inspection, while minimising the amount of erroneously flagged good
data. On average, approximately 3\,\% and 15\,\% of the data were
flagged in the lower and higher band, respectively.

After flagging, the bandpass calibration was removed as it may have
been affected by RFI in the calibrator observations, and
repeated. Phase and amplitude fluctuations throughout each observing
run were corrected using the interleaved calibrator scans, and the
amplitudes were scaled by correction factors derived from the
observations of the primary calibrator. The data were then split by
pointing and imaged.

\subsubsection{Imaging}

The data for each of the 20 pointings were imaged separately using
uniform weighting and a pixel size of $2.0''$. The 25 frequency
channels were gridded separately to increase the $(u,v)$ coverage. The
relatively high fractional bandwidth of the observations (15\,\%)
required the use of Miriad's implementation of multi-frequency clean,
MFCLEAN, for deconvolution, to account for spectral indices across the
observed bandwidth and to reduce sidelobes. After a first iteration,
model components with a flux density of more than $1\,{\rm
mJy\,beam^{-1}}$ were used in phase self-calibration, to correct
residual phase errors. The data were then re-imaged and cleaned with
5000 iterations, at which point the sidelobes of strong sources were
found to be well below the thermal noise. The models were convolved
with a Gaussian of $10.26^{\prime\prime}\times7.17^{\prime\prime}$
diameter at position angle $0^\circ$, and the residuals were
added. The restored images of the 20 pointings were merged in a linear
mosaic using the Miriad task LINMOS, which divides each image by a
model of the primary beam to account for the attenuation towards the
edges of the image, and then uses a weighted average for pixels which
are covered by more than one pointing. As a result, pixels at the
mosaic edges have a higher noise level. Regions beyond a perimeter
where the primary beam response drops below 3\,\% (this occurs at a
radius of $35.06^\prime$ from the centre of a pointing) were blanked.

Imaging of the data turned out to be challenging, but the sensitivity
of the image presented here is mostly within 25\,\% of the predicted
sensitivity. In the south-eastern corner of the mosaic, mild artefacts
remain due to the presence of the 3.8\,Jy radio source PKS\,0033-44,
which is located about $1^\circ$ away from the centre of pointing
13. The noise level of the present image could only be reached by
including this source in the CLEANed area. Because of a combination of
the high resolution of the image, the distance of the source from the
pointing centre, and the requirement of multi-frequency clean to
provide images which are three times larger than the area to clean, we
had to generate very large images with 16384 pixels on a side, plus an
additional layer of the same extent for the spectral index. These
images cannot be handled by 32-bit computers because the required
memory exceeds their address space, and we had to employ a 64-bit
machine to image the data.

The cause of the residual sidelobes is still the subject of
investigation. At present, we suspect that non-circularities in the
sidelobe pattern of the primary beams are the culprit. The interfering
source sits on the maximum of the first antenna sidelobe and, in the
course of the observations, rotates through the sidelobe pattern due
to the azimuthal mounting of the antennas. We have measured the
primary beam response of two ATCA antennas in great detail using a
geostationary satellite at 1.557\,GHz, and derived a model of their
far-field reception patterns. Unfortunately, we were unable to
reproduce the sidelobe pattern arising from PKS\,0033-44, and no
correction from this exercise has been applied to our data.

\subsubsection{Image properties}

The sensitivity is not uniform across the image due to primary beam
attenuation, however, it is quite homogeneous in the central
1\,deg$^2$ of the image. A cumulative histogram of an image of the
noise in this area, made with SExtractor (\citealt{Bertin1996}),
revealed that only 2\,\% of the image has a noise of 22\,\uJy\ or
less, consistent with the theoretical expectations. However, 75\,\% of
pixels have a noise of 27.5\,\uJy\ or less, which is 25\,\% higher
than the expected noise.  We conclude that in the regions which are
not affected by sidelobes from PKS\,0033-44 the sensitivity of the
image is close to the theoretical expectations.

\subsubsection{Clean bias}

Clean bias is an effect in deconvolution which redistributes flux from
point sources to noise peaks in the image, thereby reducing the flux
density of the real sources. As the amount of flux which is taken away
from real sources is independent of the sources' flux densities, the
fractional error this causes is largest for weak sources. The effect
of clean bias in our calibration procedure has been analysed as
follows. We have added to the data of one pointing (rms=30\,\uJy)
132 point sources at random positions, with flux densities between
150\,\uJy\ and 3\,mJy. The number of sources added with a particular
SNR were N=40 (5\,$\sigma$), 15 (6\,$\sigma$), 15 (7\,$\sigma$), 15
(8\,$\sigma$), 15 (9\,$\sigma$), 10 (10\,$\sigma$), 10 (12\,$\sigma$),
5 (16\,$\sigma$), 3 (20\,$\sigma$), 2 (30\,$\sigma$), 1
(50\,$\sigma$), and 1 (100\,$\sigma$). 

\begin{figure}
\includegraphics[width=\linewidth]{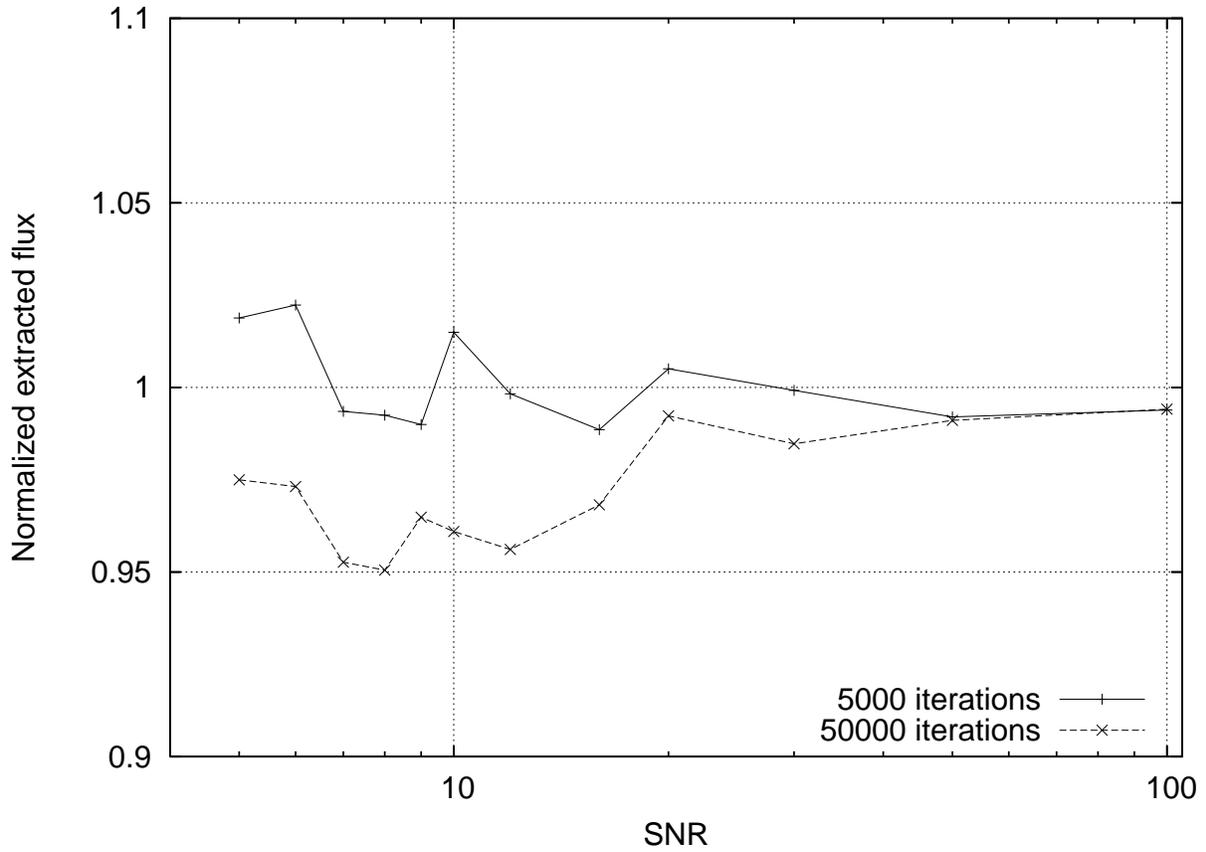}
\caption{Results from our tests for clean bias. Shown is the median
normalized flux of sources extracted from simulated images as a
function of SNR. Using 5000 iterations in cleaning does not produce a
significant clean bias, but using 50000 iteration does, although the
bias is comparatively small.}
\label{fig:clean_bias}
\end{figure}

The data have then been used to form an image in the same way as the
final image was made, and each source's flux density was extracted
using a Gaussian fit, and then divided by the injected flux
density. This test was repeated 30 times to build up significant
statistics, in particular for the sources with high SNR. We found that
using 5000 iterations in cleaning did not cause a significant clean
bias ($<2.5$\,\%), whereas using 50000 iterations did cause the
extracted fluxes to be reduced by up to 5\,\%
(Figure~\ref{fig:clean_bias}). We conclude that the flux densities in
our catalogue are only marginally affected by clean bias.

\subsubsection{Comparison to earlier observations}

We have compared the flux densities and positions of components in our
image to those of \cite{Gruppioni1999} (G99). We have obtained their
image and selected 83 isolated components with $S>0.5$\,mJy in regions
where our noise level was below 30\,\uJy. All sources were detected
with an SNR$>$6 by G99. These sources were grouped into bins with
$2^n$\,mJy to $2^{n+1}$\,mJy (n=-1,0,1,2,3,4), the flux densities were
extracted from G99's image using the same methods as for our image,
and the ratios $S/S_{\rm G99}$ were computed. The median ratios were
1.36 (0.5\,mJy-1\,mJy), 1.43 (1\,mJy-2\,mJy), 1.19 (2\,mJy-4\,mJy),
and 1.16 (4\,mJy-8\,mJy). The two highest bins with 8\,mJy-16\,mJy and
16\,mJy-32\,mJy had ratios very close to one, but only two
measurements each, hence the statistics are not reliable.

Our analysis suggests that our flux densities are systematically
higher than G99's, although $S/S_{\rm G99}$ appears to approach unity
towards higher flux densities. We have found that our flux extraction
procedure reproduces the catalogued fluxes of G99 to within 3\,\%,
hence we conclude that our procedure is working and the effect is
real. The cause of this discrepancy is not known, but possible
explanations are (i) calibration differences: G99 used amplitude
self-calibration with a relatively sparse array and very short
solution intervals, which may have affected the flux densities. We did
not use amplitude self-calibration at all because it was not found to
improve our image significantly; (ii) $(u,v)$ coverage: G99 had only
one configuration at the ATCA whereas we had six, yielding more
constraints in deconvolution. Also G99 imaged the data from both IF
bands separately and averaged the images later, thus using only one
half of their data in the deconvolution stage.

We also tested for a systematic position offset between the components
of G99 and ours. We found a mean offset of
$0.112^{\prime\prime}\pm0.016^{\prime\prime}$ in right ascension and
of $0.017^{\prime\prime}\pm0.022^{\prime\prime}$ in declination, and
conclude that systematic position offsets are negligible.

\subsection{{\it Spitzer} observations}

The {\it Spitzer} observations of the ELAIS-S1 field were carried out as
part of the {\it Spitzer} Wide-Area Infra-Red Extragalactic Survey (SWIRE)
program, as described by \cite{Lonsdale2003}. Approximately
6.9\,deg$^2$ were observed in the ELAIS-S1 region at 3.6\,\um,
4.5\,\um, 5.8\,\um, and 8.0\,\um\ with the Infrared Array Camera (IRAC)
and at 24\,\um\ with the Multiband Imaging Photometer (MIPS). The
sensitivities in the five bands are 4.1\,\uJy, 8.5\,\uJy, 48.2\,\uJy,
53.0\,\uJy, and 252\,\uJy. Here we use the fourth data release,
containing more than 400.000 sources (Surace et al. 2007, in prep.).

\subsection{Optical observations}

The optical follow-up observations of the ELAIS-S1 field are called
the ESO-Spitzer Imaging Extragalactic Survey (ESIS). The observations
were carried out with the Wide Field Imager (WFI) of the 2.2\,m
La~Silla ESO-MPI telescope and with the VIsible Multi Object
Spectrograph (VIMOS) on the VLT, to cover 5\,deg$^2$ in BVRIz. Only
approximately 1.5\,deg$^2$ have yet been covered (\citealt{Berta2006})
with WFI and these data are included in our catalogue. The filters
used are WFI B/99 (later replaced by B/123), V/89 and Rc/162, and the
catalogue is 95\,\% complete at 25$^{\rm m}$ in the B and V bands, and
at 24.5$^{\rm m}$ in the R band (all in Vega units).

\section{Image analysis}
\label{sec:images}

\subsection{Component extraction}

This section describes the procedure we used to extract radio sources
from the image and to subsequently match these radio sources to
infrared sources. In our terminology, a radio component is a region of
radio emission which is best described by a Gaussian. Close radio
doubles are very likely to be best represented by two Gaussians and
are therefore deemed to consist of two components. Single or multiple
components are called a radio source if they are deemed to belong to
the same object.

The rms of the image varies from 22\,\uJy\ in the best regions to
1\,mJy towards the edges of the image, caused by primary beam
attenuation. It is therefore not possible to use the same cutoff, in
terms of flux density per pixel, above which a pixel is deemed a
detection of a source and below which pixels are deemed
noise. Furthermore, flux densities measured towards the image edges
are increasingly affected by uncertainties in the primary beam model,
and we therefore restricted our image analysis to those sources which
lie in regions where the theoretical sensitivity is below 250\,\uJy.

We used SExtractor to create an image of the noise, by which we
divided the radio image to obtain an image of signal-to-noise (called
the SNR map). The SNR map has unity noise everywhere, and can be
analysed using a single criterion. We used the Miriad task IMSAD to
look for islands of SNR$>$5, and then used this catalogue as input for
a visual inspection of the total intensity image at the locations
where SNR$>$5. Sources were re-fitted using the total intensity image,
and were subsequently cross-identified with IR sources and
classified. If either of the two axes of a fitted Gaussian was smaller
than the restoring beam's corresponding axis, the fit was repeated
using a Gaussian with the major and minor axis fixed to the restoring
beam and the position angle set to zero. Also very weak sources were
in general found to be better represented with fixed-size Gaussians.

The integrated flux densities of extended sources were obtained by
integrating over the source area, rather than summing the flux
densities of their constituents. This is because even multiple
Gaussians are seldom a proper representation of extended sources, and,
using this technique, even very faint emission between components is
included.

We estimated the error of the integrated flux densities using Eq.~(1)
in \cite{Schinnerer2004}, which is based on \cite{Condon1997},
assuming a relative error of the flux calibration of 5\,\% whereas
\cite{Schinnerer2004} assumed 1\,\%. In the case of extended sources,
where the integrated flux density was measured by integrating over a
polygon in the image, we assumed a 5\,\% scaling error and added to
that in quadrature an empirical error arising from the shape and size
of the area over which was integrated:

\begin{equation}
\Delta S = \sqrt{(0.05S)^2 + (10^{-7}/S)^2}
\end{equation}

where $S$ is the flux density in Jy. For extended sources with 10\,mJy,
1\,mJy and 0.5\,mJy, the total errors are thus 0.5\,mJy (5\,\%),
0.11\,mJy (11\,\%), and 0.2\,mJy (40\,\%), respectively, which
describe the errors found empirically reasonably well.

The uncertainties in the peak flux densities were estimated using
Eq.~(21) in \cite{Condon1997}. Errors in right ascension and
declination are the formal errors from Gaussian fits plus a
$0.1^{\prime\prime}$ uncertainty from the calibrator position added in
quadrature.

\subsubsection{Deconvolution of components from the restoring beam}

All radio components were deconvolved from the restoring beam. If a
deconvolution was not possible, or the deconvolution yielded a point
source, the component was deemed to be unresolved and the deconvolved
size has been left blank in Table~\ref{tab:radio1}.

\subsection{The cross-identification process}

The cross-matching process was as follows. The region used for the fit
and the ellipse indicating the FWHM were inspected, along with the
corresponding parts of the following images: the SNR map, a naturally
weighted radio image with lower resolution (and slightly higher
sensitivity), a superuniformly weighted radio image with higher
resolution (but lower sensitivity), and the 3.6\,\um\ SWIRE image with
superimposed SNR map contours. Furthermore, the locations of
catalogued SWIRE sources within $30^{\prime\prime}$ of the fitted
coordinates were shown on the SWIRE images.

It was then decided (i) whether each radio component was a genuine
detection or likely to be a sidelobe, (ii) how it could be matched to
catalogued or uncatalogued SWIRE sources, (iii) whether multiple radio
components constituted radio emission from a single object, and (iv)
whether extended components needed to be divided into
sub-components. Emission deemed to be sidelobes was found
predominantly towards the edges of the image and associated with, and
directly adjacent to, strong sources.

Most sidelobes were discovered because the naturally-weighted image,
which has a different sidelobe pattern and higher sensitivity but
lower resolution, showed no evidence of a source at the position of a
possible source in the uniformly weighted image. Our catalogue of
radio components contains 1366 components; 15 were deemed to be
sidelobes and have been marked as such (all with SNR$<$6), leaving 1351
genuine radio components.

\begin{figure}
\includegraphics[width=\linewidth]{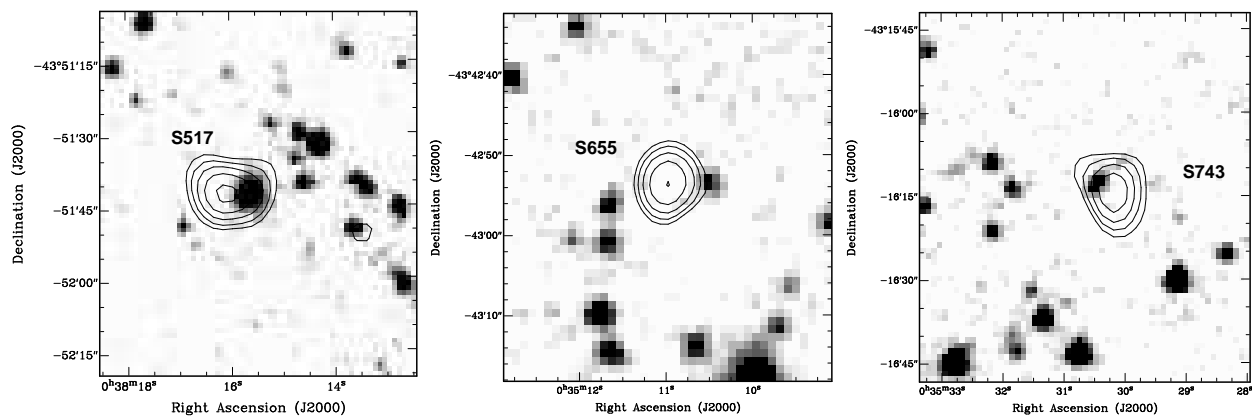}
\caption{Examples of sources with relatively large ($>3''$)
separations between the fitted radio position and the catalogued SWIRE
position. Shown are the radio SNR contours from SNR=4 and increasing
by factors of $\sqrt{2}$, superimposed on the SWIRE 3.6\,\um\ image as
greyscale. {\it Left:} S517 is strong and clearly extended towards
SWIRE4\_J003815.62-435142.0, which was deemed to be associated despite
a separation of $4.4''$. {\it Middle:} Source S655 is separated by
$5.3''$ from its SWIRE counterpart. Shown here is a portion of the
radio image made with super-uniform weighting, and hence higher
resolution, which shows the extension clearly. {\it Right:} Source
S1034 is similarly extended towards a SWIRE source, with a separation
of $3.4''$.}
\label{fig:large_sep}
\end{figure}

The separation between a radio component and a SWIRE source cannot
easily be used as a parameter in the cross-identification process. In
some cases, despite a relatively large separation, the
cross-identification is relatively clear because the SWIRE source is
extended towards the SWIRE source, such as in the examples shown in
Figure~\ref{fig:large_sep}.

1134 radio components (88.9\,\%) could be characterized properly by a
single Gaussian and were judged to be the only radio counterpart of a
catalogued SWIRE source. A fraction of these displayed the morphology
of doubles in a superuniformly weighted image. 15 sources (1.2\,\%)
had uncatalogued SWIRE counterparts.

32 sources (2.5\,\%) were deemed to be radio doubles, consisting of
two radio components; and 26 sources (2.0\,\%) consisted of two or
more components, displaying more complex morphologies like triplets or
core-jet morphologies.

We have tested for systematic radio-IR position offsets by calculating
the average offsets of 533 sources which consist of a single radio
component and a catalogued SWIRE counterpart, and have SNR$>$10. The
offsets have a mean of $(0.08\pm0.03)''$ in right ascension and
$(0.06\pm0.03)''$ in declination. Although the offset is formally
significantly different from zero, we note that it is less than a
tenth of a pixel in the radio image.

\begin{figure}
\includegraphics[width=0.9\linewidth]{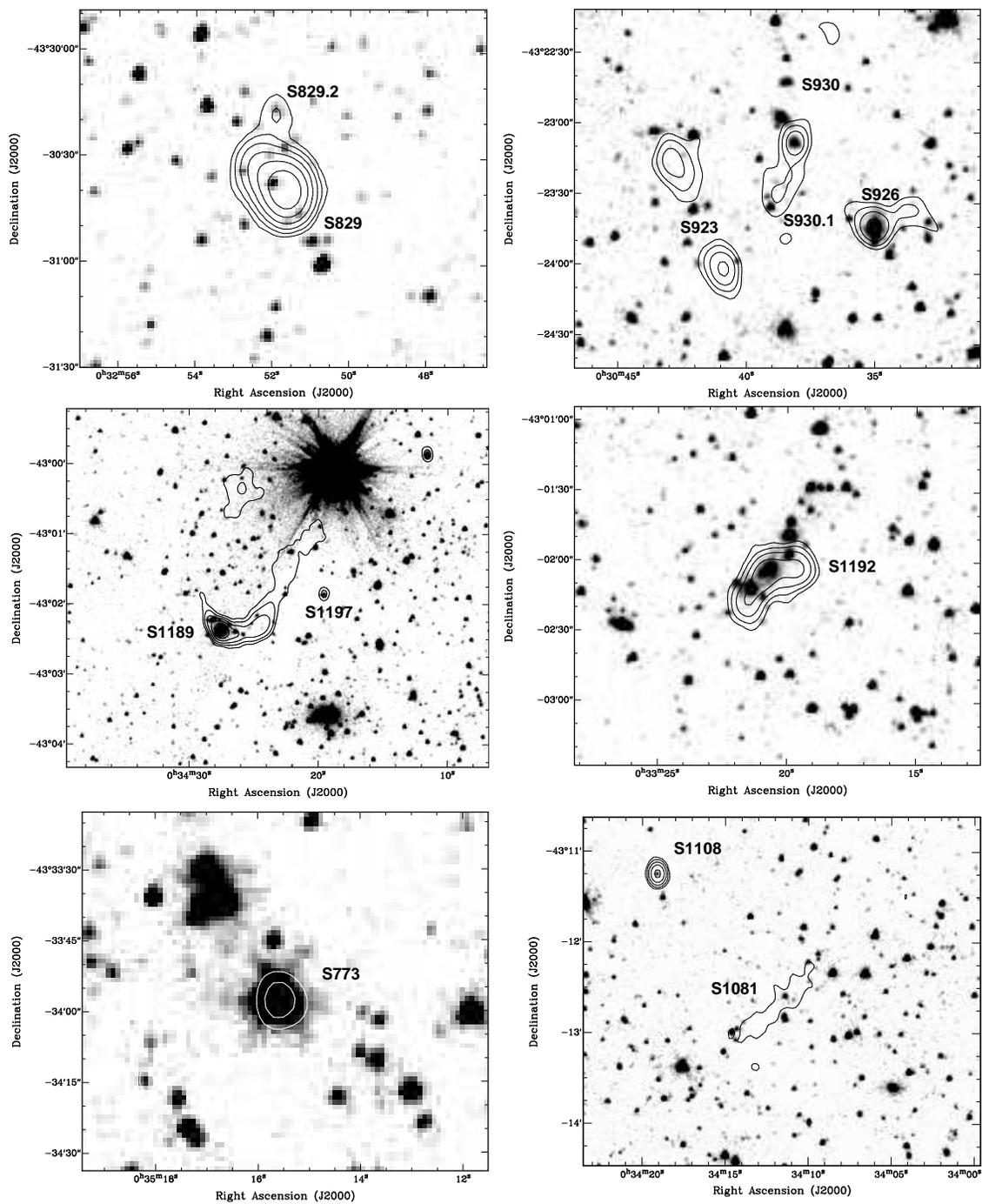}
\caption{\small Six sample extracts from the radio image (contours),
superimposed on the 3.6\,\um\ {\it Spitzer} image
(greyscale). Contours start at SNR=4 and increase by factors of
two. The rms in the images is 27\,\uJy\ (top left), 46\,\uJy\ (top
right), 49\,\uJy\ (middle left), 43\,\uJy\ (middle right), 29\,\uJy\
(lower left), and 28\,\uJy\ (lower right). See the text for a
detailed description of these sources.}
\label{fig:examples}
\end{figure}

All sources classified as radio doubles have been reviewed using the
criteria developed by \cite{Magliocchetti1998} based on an analysis of
the FIRST survey (\citealt{Becker1995}): Two radio components are
likely to be part of a double when (a) their separation measured in
arcsec is less than $100(S/100)^{0.5}$, where S is the total flux of
the two constituents, and (b) their flux densities do not differ by
more than a factor of four. We give the results of this test in the
source table. It should be noted that the test has been derived from a
large sample of galaxies (236000) and is purely
empirical. Furthermore, the FIRST survey is shallower (rms=0.14\,mJy)
than ours, and so statistically may contain different objects from the
survey presented here. It is therefore no surprise that some of our
radio sources which are clearly radio doubles fail the test. For
example, S923 (Figure~\ref{fig:examples}) fails on criterion (a), but
satisfies criterion (b).

\subsection{The false cross-identification rate}

Because the SWIRE field has a high IR source density (58700 sources
per deg$^2$), there is some chance that a radio component falls within
a few arcseconds of an infrared source, although it is not physically
connected to it. The two sources would be wrongly cross-matched, and
hence there is a fraction of erroneous cross-identifications in our
source catalogue, an upper limit of which we estimate as follows.

From the source density, one can calculate that on average 0.01423
SWIRE sources fall within $1''$ of any one point in the field. The
number of SWIRE sources within $1''-2''$ of any one point is 0.0427,
and within $2''-3''$ is 0.0711. We have confirmed these numbers
experimentally by searching near several hundred random positions in
the SWIRE catalogue.

In our catalogue, 1134 sources consist of a single component and have
a good SWIRE cross-identification. Of these, 656 have a separation of
less than $1''$, 350 have a separation of $1''-2''$, 86 have a
separation of $2''-3''$, and 45 have a separation of more than $3''$.

Of the original 1134 cross-identifications, a fraction of 0.01423, or 16
sources, are expected to be purely coincidental, and are found among
the 656 sources with sub-arcsec cross-identifications. Thus, a
fraction of $16/656=0.024$ is likely to be coincidental (and wrong).

With the sub-arcsec cross-identifications now accounted for,
$(1134-656)=478$ sources remain. Of these, a fraction of 0.0427, or 20
sources, will fall within $1''-2''$ of an infrared source by
coincidence. Thus, a fraction of $20/481=0.042$ is coincidental.

Repeating the steps above leaves $(1134-656-350)=128$ sources which have
not yet been cross-identified. Putting 128 sources randomly on the
SWIRE image yields a coincidental counterpart within $2''-3''$ for a
fraction of 0.0711, or 9 sources. Thus, a fraction of $9/86=0.105$ is
coincidental. The statistics of the remaining 45 sources with
separations $>3''$ are not meaningful because the separations are
dominated by extended radio objects which are not expected to coincide
with infrared sources. A summary of this estimate is shown in
Table~\ref{tab:fxids}.

We stress that the rates of false cross-identifications given here are
upper limits. A false cross-identification does not only require a
false counterpart within a few arcseconds of the radio position, but
it also requires that the true counterpart is much fainter than the
false one. The second requirement reduces the rate of false
cross-identifications well below our estimate.

\section{The component and source catalogues}
\label{sec:cats}

Following \cite{Norris2006a} we publish two catalogues, one containing
the component data (Table~\ref{tab:radio1}), and one containing radio
sources and their infrared counterparts (Table~\ref{tab:source1}).

\subsection{The components catalogue}

The component catalogue contains information about Gaussian components
fitted to the radio image. It does not contain information about the
grouping of components to sources, which is exclusively left to the
source catalogue in the next section.

\subsection{The source catalogue}

\begin{figure}
\includegraphics[width=\linewidth]{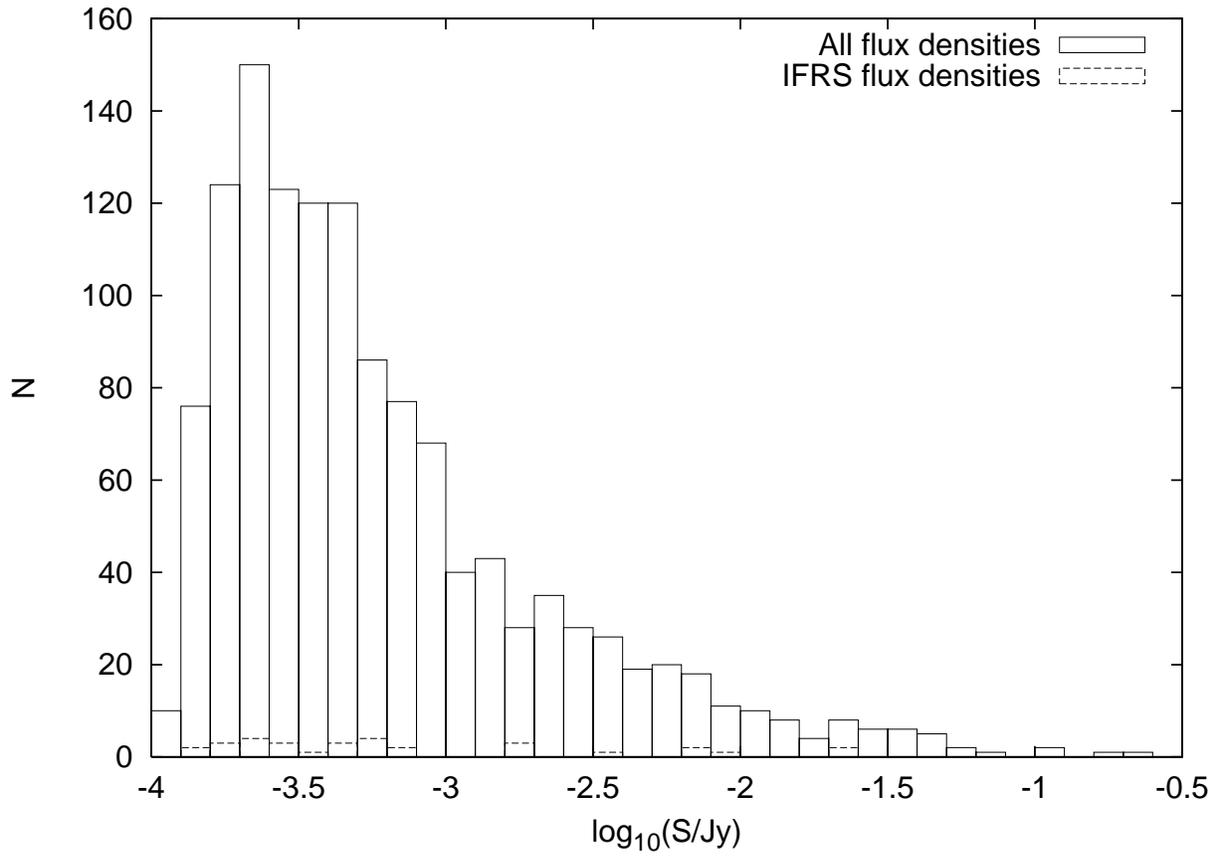}
\caption{A histogram of the integrated flux densities of the sources in
our survey. A histogram of the IFRS flux densities is drawn with
dashed lines.}
\label{fig:fluxes}
\end{figure}

The distribution of integrated flux densities for the 1276 catalogued
sources is shown in Figure~\ref{fig:fluxes}. We have carried out a
Kolmogorov-Smirnov-test using the ELAIS-S1 and CDFS integrated flux
densities, to test the likelihood that the two samples are drawn from
the same parent distribution. Because the two fields have different
sensitivities, the catalogues cannot be compared in full, but a flux
cutoff has to be used. Furthermore, we restricted the test to sources
within $48'$ of the field centres and required an rms of between
30\,\uJy\ to 40\,\uJy, to exclude regions with elevated noise levels
towards the image edges. We find that when only sources with flux
densities of more than 0.5\,mJy are compared (ELAIS-S1: 137 sources,
CDFS: 130 sources), the probability that the two samples are drawn
from the same parent distribution is 73.7\,\%. When the minimum
required flux density is lowered to 0.4\,mJy (ELAIS-S1: 179 sources,
CDFS: 151 sources) or 0.3\,mJy (ELAIS-S1: 222 sources, CDFS: 186
sources), the probabilities are 18.3\,\% and 25.8\,\%,
respectively. We conclude that in regions with similar sensitivities
the distribution of radio sources in the ELAIS-S1 and CDFS fields is
identical at a flux density level of more than 0.3\,mJy.

In the source catalogue, comments on the cross-match and the radio
morphology are recorded as follows. If no comment is given, we had no
doubt about the identification; "uncatalogued counterpart" means that
we had no doubt that the radio source is associated with a clearly
visible IR source at either 3.6\,\um\ or 24\,\um\ which is not listed
in the SWIRE catalogue (data release 4); "IFRS" means that a radio
source could not be reasonably matched to any IR counterpart at all
and did not appear to be associated with another radio source;
"confused XID" means that the radio source is likely to be associated
with the SWIRE source we give, but that other sources cannot be ruled
out; "unclear XID" means that the identification was too ambiguous to
make a reasonable choice. We also comment on the radio morphology if
the source is anything but a single Gaussian. In the case of
multiple-component sources we give the component numbers which were
deemed to be associated with the source, and we comment on extension
or blending with other radio sources. The coordinates of sources are
generally those of the radio observations, but in the case of sources
with more than one component and with a clear IR counterpart, the
SWIRE coordinates have been adopted as the source position. In the
case of more than one component without a clear IR component the
flux-weighted mean of the radio components has been used.

\subsection{Identification of sources with other catalogues and literature data}

The ELAIS-S1 region has already been surveyed with the ATCA at
1.4\,GHz by \cite{Gruppioni1999} with a $1\,\sigma$ sensitivity of
80\,\uJy, and we have cross-matched their catalogue to ours, resulting
in 366 matches. We have also searched the NASA Extragalactic
Database\footnote{\url{http://nedwww.ipac.caltech.edu/index.html}} for
objects within $2^{\prime\prime}$ of the sources in our catalogue, and
found matches to 105 sources, sometimes with multiple names. We mostly
give the designations from the ELAIS 15\,\um\ catalogue
(\citealt{Oliver2000}), the APMUKS catalogue (\citealt{Maddox1990}),
and the 2MASS catalogue (\citealt{Skrutskie2006}). These
cross-identifications have been included in Table~\ref{tab:source1}.

\begin{figure}
\includegraphics[width=\linewidth]{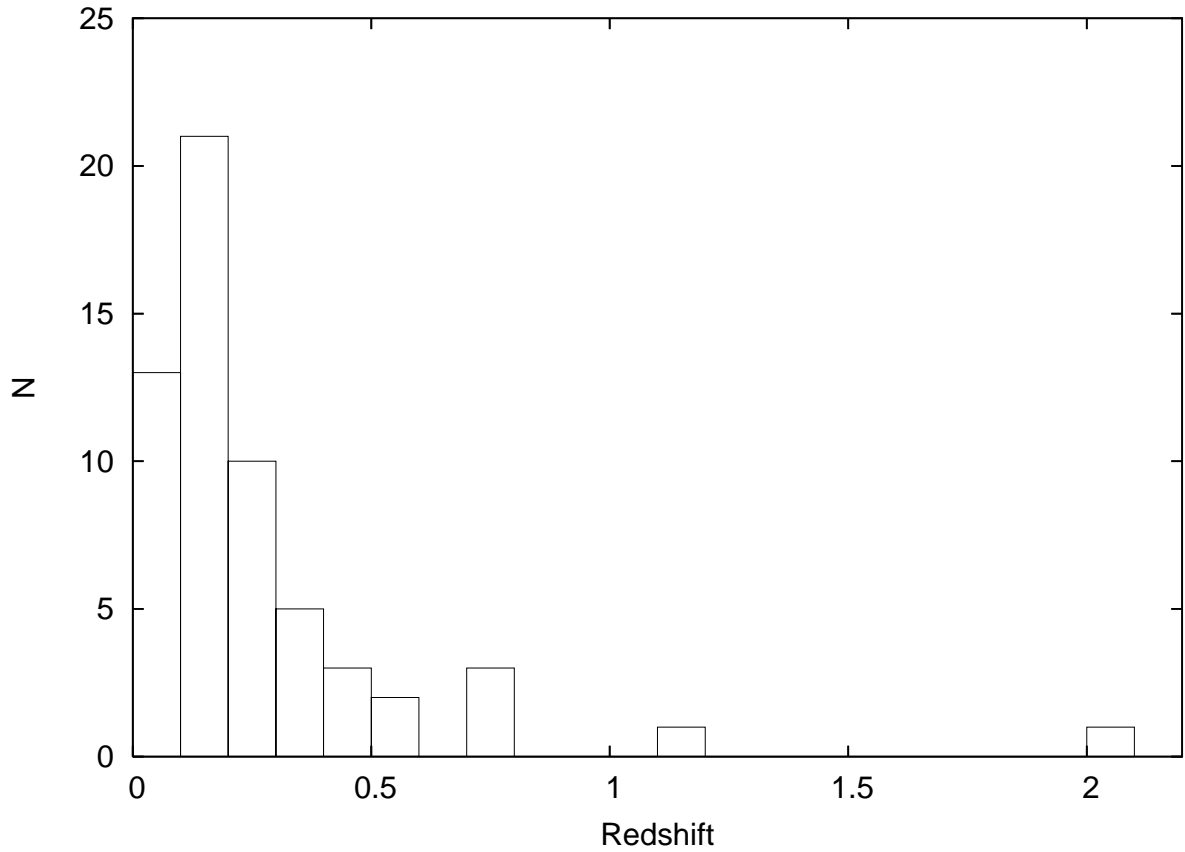}
\caption{A histogram of the 59 redshifts available for objects in our
catalogue, taken from the literature. There is no hint of large-scale
structure, but this may be hidden by too few data points.}
\label{fig:redshifts}
\end{figure}

We searched for available redshifts and found that 59 objects within
$2^{\prime\prime}$ of our sources had catalogued redshifts, mostly
from \cite{LaFranca2004}. A histogram of the redshifts is shown in
Figure~\ref{fig:redshifts}. Unlike in the CDFS, there is no indication
of cosmic large-scale structure in this histogram. However, the number
of redshifts is small and may not be sufficient to show inconspicuous
large-scale structure.

We have cross-matched our source catalogue to sources from the Sydney
University Molonglo Sky Survey (SUMSS, \citealt{Bock1999}), which is a
survey of the southern sky at 843\,MHz, using the Molonglo Observatory
Synthesis Telescope (MOST). The sensitivity of SUMSS is of the order
of $\sim1\,{\rm mJy\,beam^{-1}}$, so that the faintest sources have a
flux density of the order of $\sim5\,{\rm mJy}$. Assuming a spectral
index of $\alpha=-0.7$ ($S\propto\alpha$), typical for radio emission
from AGN, this corresponds to $S_{\rm 1.4GHz}=3.5\,{\rm mJy}$, so only
the brightest ATLAS sources will be present in SUMSS. We found 73
matches to sources catalogued in the 1 June 2006 data
release\footnote{\url{http://www.astrop.physics.usyd.edu.au/sumsscat/}}
and give the results in Table~\ref{tab:sumss}. There were no SUMSS
sources without 1.4\,GHz counterpart in the ATCA image.

We have also searched for counterparts in the AT20G survey
(\citealt{Ricci2004}), which is a survey of the southern sky with the
ATCA at 18\,GHz, but found no match.

\section{Classification}
\label{sec:class}

\subsection{AGN}

Here we discuss the classification of sources as AGN based on their
morphology, their ratio of 24\,\um\ to radio flux, and using
literature information.

Radio sources exhibiting a double-lobed, triple, or more complex
structures, e.g. with jets, were generally classified as AGN. Examples
for classification based on morphology are S829, S923, S926, S930.1,
S1192 and S1189, all of which are described in more detail in
Section~\ref{sec:individual}.

From {\it Spitzer} 24\,\um\ and VLA 20\,cm detections in the First
Look Survey (\citealt{Condon2003}), \cite{Appleton2004} derive
$q_{24}={\rm log}(S_{24\um}/S_{20{\rm cm}})=0.84$. Here, sources with
${\rm log}(S_{24\um}/S_{20{\rm cm}})<-0.16$, i.e. more than 10 times
the radio flux density as predicted by the radio-infrared relation,
were classified as AGN. 

\begin{figure}
\includegraphics[width=\linewidth]{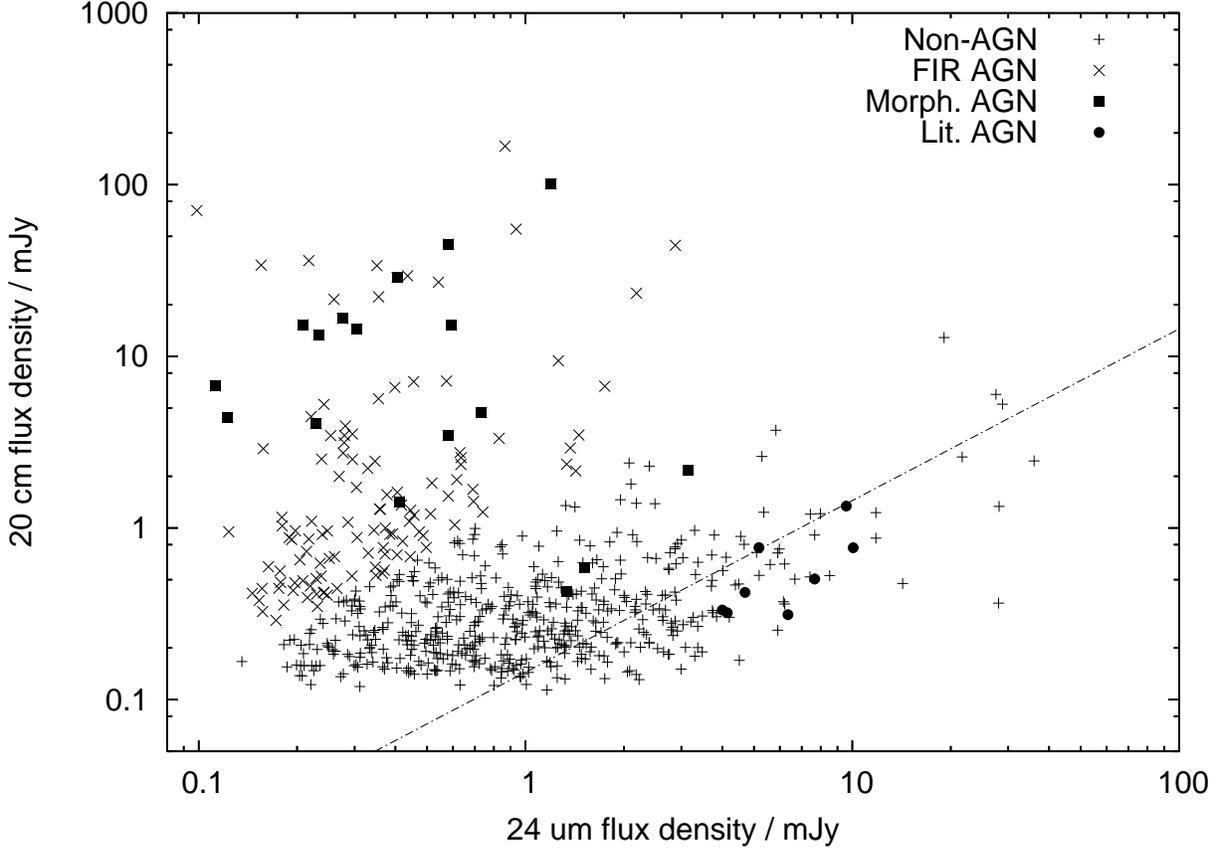}
\caption{20\,cm vs. 24\,\um\ flux densities of all sources, with AGN
plotted separately. Symbols indicate the type of AGN classification:
pluses show non-AGN; crosses indicate AGN classified based on an
10-fold excess of radio emission compared to the infrared-radio
emission derived by \cite{Appleton2004}; filled squares indicate AGN
classified based on their radio morphology; and filled circles
indicate sources classified as AGN in the literature.  The line
indicates $q_{24}={\rm log}(S_{24\um}/S_{\rm 20cm})=0.84$ found by
\cite{Appleton2004}. The flattening of the distribution towards lower
values of $S_{24\um}$ is caused by the limited sensitivity of the
radio observations, which at low 24\,\um\ flux densities are only able
to pick up objects with comparatively high radio flux densities. For
a detailed analysis of the radio-infrared relation derived from the
ELAIS-S1 radio and 24\,\um\ data, see Boyle et al. (2007).}
\label{fig:radio-ir}
\end{figure}

In total 75 sources were classified as AGN based on their radio
morphology, 128 sources based on their radio excess compared to the
radio-infrared relation at 24\,\um, and 9 sources had been classified
as AGN by \cite{LaFranca2004}, based on optical spectroscopy. 14
sources were classified as AGN using more than one criterion, and thus
198 sources were classified as AGN. We note that, with the exception
of the three sources S606, S717, and S813, all sources which were
classified as AGN based on their morphology {\it and} which had
catalogued 24\,\um\ flux densities were also classified as AGN based
on their departure from the radio-infrared relation as given by
\cite{Appleton2004}. We plot the 20\,cm flux densities as a function
of 24\,\um\ flux densities of all sources in
Figure~\ref{fig:radio-ir}. AGN are plotted separately according to how
they have been classified.

\subsection{Infrared-Faint Radio Sources (IFRS)}

We find 31 sources with no detectable infrared counterpart. These
sources have been dubbed ``Infrared-Faint Radio Sources'', or IFRS, by
\cite{Norris2006a}, and may be more extreme cases of the ``Optically
Invisible Radio Sources'' found by \cite{Higdon2005}. As they are
invisible in the optical and infrared, there is only very limited
information available. A Kolmogorov-Smirnov test reveals a 80.1\,\%
probability that the distribution of flux densities of the IFRS is
drawn from the same parent distribution as all flux densities, though
the IFRS sources tend to have lower radio flux densities than the
entire sample. We have carried out VLBI observations of three IFRS in
our sample (S427, $S_{\rm 1.4GHz}=21.4\,{\rm mJy}$; S509, $S_{\rm
1.4GHz}=22.2\,{\rm mJy}$; and S775, $S_{\rm 1.4GHz}=3.6\,{\rm mJy}$)
to determine whether they are AGN hosts and the contribution to the
arcsec-scale flux density from an AGN, but the results are not yet
available. However, \cite{Norris2007b} have successfully detected an
IFRS in the CDF-S field.

\section{Notes on individual sources}
\label{sec:individual}

We comment on a few examples to illustrate the cross-identification
process. The sources discussed here are shown in
Figure~\ref{fig:examples}.

\begin{figure}
\includegraphics[width=\linewidth]{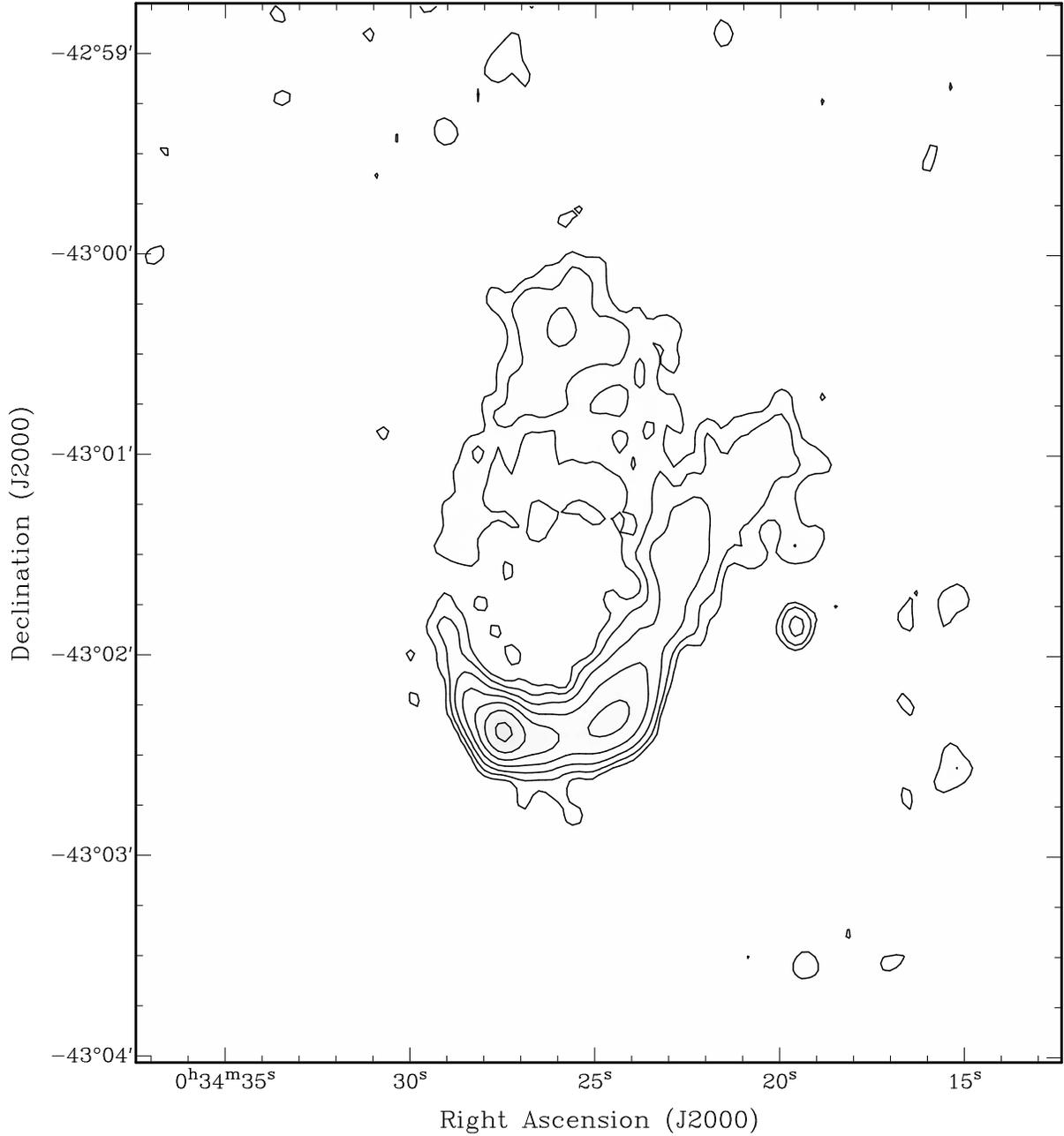}
\caption{Source S1189, drawn with contours starting at
$2\,\sigma=70\,\uJy$ and increasing by factors of two. The two jets
are strongly bent backwards, and their far ends almost touch each
other. The morphology suggests that the source is moving through a
relatively dense medium, indicating the presence of a yet unknown
galaxy cluster.}
\label{fig:1189}
\end{figure}

\begin{itemize}
\item{\it Sources S829 and S829.2} S829 is an example of a mildly
extended object, which is best represented with two Gaussians (C829
and C829.1). However, at higher resolution it begins to resemble a
double-lobed or core-jet morphology, and it is centred on the IR
source SWIRE4\_J003251.97-433037.2 in between the two radio
components, and hence was classified as an AGN. The nearby source
S829.2 is a relatively weak radio source (0.30\,mJy) which coincides
($\theta \sim 1.5^{\prime\prime}$) with SWIRE4\_J003251.87-433016.7.

\item{\it Sources S923, S930, S930.1 and S926} These four sources lie within
less than $2^\prime$ of each other and form a striking quartet at
first sight. Source S923 is without doubt a classical double-lobed
radio galaxy with an integrated flux density of 5.9\,mJy. The SWIRE
source SWIRE4\_J003042.10-432335.4 is located on the line connecting
the peaks of the two constituent radio components C923 and C931 and is
therefore identified as the infrared counterpart. Source S930 is an
otherwise inconspicuous radio source with an infrared counterpart,
SWIRE4\_J003038.21-432305.9, within $0.28^{\prime\prime}$. The
naturally weighted image indicates a faint bridge of emission between
components C930 and C941, hence both components have been grouped into
S930.  It blends with source S930.1, which consists of the two faint
radio components C930.1 and C930.2 with integrated flux densities of
0.47\,mJy and 0.55\,mJy, respectively. In between components C930.1
and C930.2 is a very faint, uncatalogued infrared source, and thus the
radio morphology together with the location of the IR source indicates
that this is a double-lobed radio galaxy. Source S655 has a relatively
bright IR counterpart (SWIRE4\_J003035.03-432341.6) centred on the
brighter one of its two radio components C926 and C926.1, with
2.7\,mJy and 0.93\,mJy, respectively. Unlike in sources S923 and
S930.1, the IR source is centred on one of its constituents, but it
was deemed more likely that both C926 and C926.1 are associated with
this source rather than to postulate that C926 is the radio
counterpart to SWIRE4\_J003035.03-432341.6 and that C926.1 is a
separate source with no IR counterpart.

\item{\it Sources S1189 and S1197} Source S1189 is a beautifully extended,
large radio source. The number of constituent radio components is
somewhat arbitrary, but there exists a low-SNR bridge of emission
which connects the main part of the source and component C1212,
$2^\prime$ north, as well as many more low-SNR patches in between. The
brightest part of S1189 is centred on SWIRE4\_J003427.54-430222.5
(separation $0.75''$), which we therefore identify as IR counterpart,
and which has the morphology of an elliptical galaxy in optical
images. Source S1197, $0.60^{\prime\prime}$ from
SWIRE4\_J003419.55-430151.7, is unlikely to be connected to
S1189. S1189 has the shape and extent of Wide-Angle Tail galaxies
(WAT, \citealt{Miley1980}) such as NGC\,1265
(\citealt{Owen1978}). Their characteristic C-shape is believed to be
caused by ram pressure against the jets while the galaxy is moving
through the intracluster medium. The jets in S1189 are strongly bent
backwards and almost touch another at the far ends. This is
illustrated in Figure~\ref{fig:1189}, where we have drawn contours
beginning at $2\,\sigma$ to emphasize the effect. WAT radio sources
can be used as cluster signposts (e.g.,
\citealt{Blanton2003}), but there is no known cluster at the position
of S1189, although there is a little overdensity of galaxies at
$115''$ to its south-west, centred on a bright galaxy with elliptical
morphology. The source is in the SUMSS catalogue with a flux density
of 36.2\,mJy, compared to a 1.4\,GHz flux density of
45.0\,mJy. However, it is clearly extended in the SUMSS image and not
well represented by a Gaussian. Integrating over the source area in
the SUMSS image yielded a total flux density of 51\,mJy, and hence a
spectral index of $\alpha=-0.25$.

\item{\it Source S1192} This source is an example of a triple radio
galaxy. It consists of the three components C1192, C1192.1, and
C1192.2 with mJy flux densities. The brightest component, C1192.1, is
centred on the IR source SWIRE4\_J003320.68-430203.6. The other two
components are several arcsec away from the nearest IR sources, and
the overall morphology thus indicates that this is a bent triple radio
galaxy. It therefore also could be a WAT.

\item{\it Source S773} Source S773 is a rather faint radio
source with $S_{\rm 20cm}=0.37\,{\rm mJy}$, but it has a very bright
infrared counterpart with $S_{24\um}=28\,{\rm mJy}$ within $0.72''$,
and is one of the few objects clearly visible in the SWIRE 70\,\um\
image. Its unusual ratio of ${\rm log}_{10}(S_{24\um}/S_{\rm
20cm})=1.89$ lets it clearly stand out in Figure~\ref{fig:radio-ir} as
a separated dot in the bottom right corner. It has been classified by
\cite{LaFranca2004} as a type 1 AGN (based on optical line widths in
excess of 1200\,km/s) at redshift 0.143. Furthermore, it is one of the
brightest X-ray sources found by
\cite{Alexander2001} in a {\it BeppoSAX} survey of the ELAIS-S1
region.

\item{\it Source S1081} Source S1081 is a very extended ($B_{\rm
maj}=93''$), low-surface brightness source. Nevertheless, its
integrated flux density is 2.4\,mJy, and there is no obvious
association with any one of the many nearby infrared sources. It is
unlikely to be a sidelobe, as this region of the image is very good
and free of artefacts. Furthermore, its extent indicates that it is
not a noise spike, which would have a similar size as the restoring
beam. We have convolved the radio image of \cite{Gruppioni1999} with a
$1'$ restoring beam, to increase its sensitivity to extended
structures, but their image was not sensitive enough to confirm or
refute the reality of S1081. The nature of this object is unclear:
given its size and the lack of a strong component it could be a
cluster radio relic. Such objects are interpreted as leftovers of
cluster mergers.

\end{itemize}

\section{The radio-infrared relation}
\label{sec:r-ir}

One of the goals of the ATLAS project is to trace the radio-infrared
relation to very faint flux levels, to determine whether the relation
exists in the early universe. Using {\it Spitzer} and VLA observations
of the First Look Survey (\citealt{Condon2003}), \cite{Appleton2004}
have determined a value of $q_{24}={\rm log}(S_{24\um}/S_{\rm
20cm})=0.84\pm0.23$. Here, we note that \cite{Boyle2007} have employed
a statistical analysis of the ELAIS-S1 radio image at the known
positions of SWIRE sources. They find $q_{24}=1.46$ using the
observations presented here, and exactly the same value of $q_{24}$
using the CDFS observations of \cite{Norris2006a}. \cite{Boyle2007}
present an extensive description of the analysis and of simulations,
and we refer the reader to their paper for details. We note, however,
that the discrepancy of the value of $q_{24}$ found by
\cite{Appleton2004} and \cite{Boyle2007} remains unresolved.

\begin{figure}
\center
\includegraphics[width=14cm]{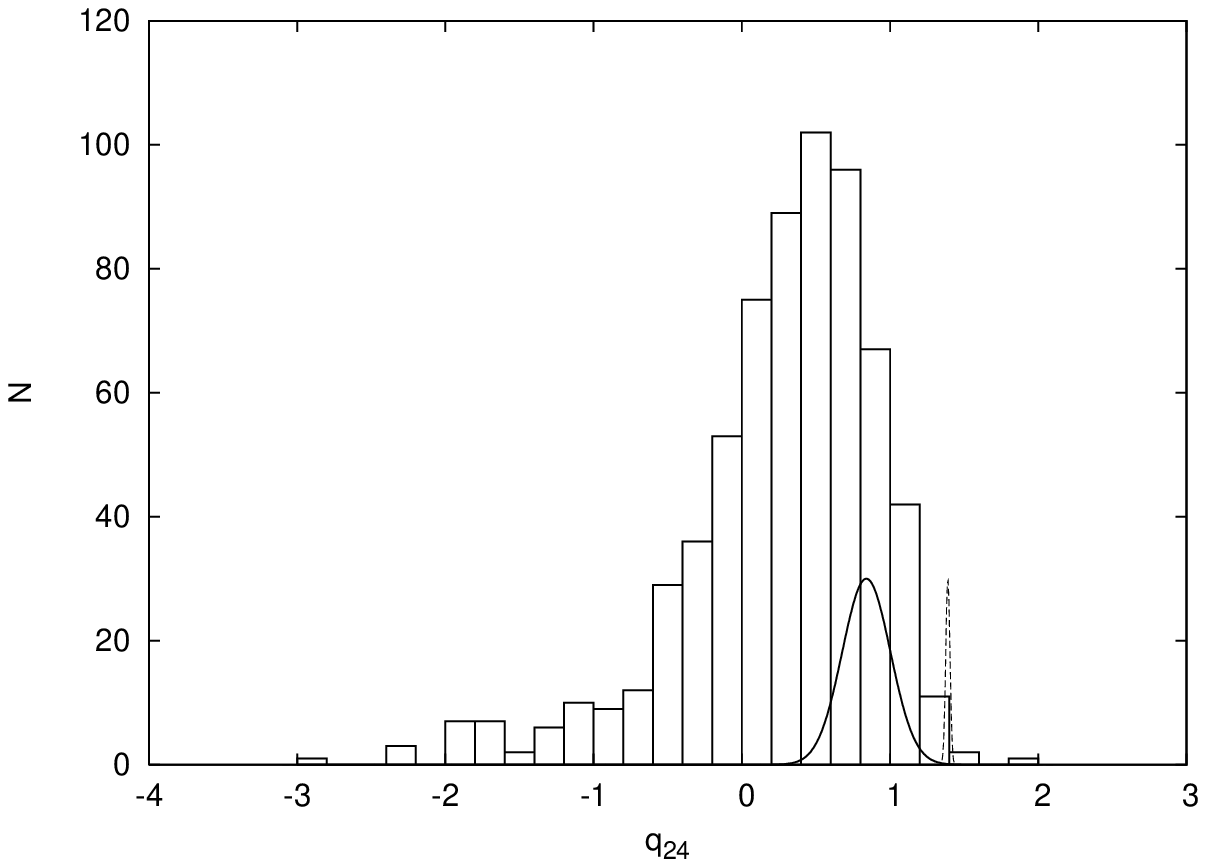}\\
\caption{Histogram of $q_{24}$ from all ELAIS-S1
data. The solid Gaussian indicates $q_{24}=0.84\pm0.23$ as found by
\cite{Appleton2004}, and the dashed Gaussian $q_{24}=1.39\pm0.02$ as
found by \cite{Boyle2007}. The histogram peak is in broad agreement
with the \cite{Appleton2004} results, and the tail towards low values
of $q_{24}$ is caused by AGN, which are included in our data but were
discarded by \cite{Appleton2004}. Why the \cite{Boyle2007} peak does
not agree with the histogram and the \cite{Appleton2004} distribution
is not understood.}
\label{fig:q24}
\end{figure}

We plot in Figure~\ref{fig:q24} a histogram of all individual values
of $q_{24}$ where 24\,\um\ fluxes were available, without
k-correction. We also indicate on the diagram the distribution (also
without k-correction) found by \cite{Appleton2004}. We note that these
authors also presented a k-correction for their data, but it was too
small to reconcile their result with the result by
\cite{Boyle2007}. The tail towards lower values of $q_{24}$ can be
explained as arising from AGN, which have a radio excess and so do not
obey the radio-infrared relation. Conversely, the sharp cutoff of the
histogram at $q_{24}$ is caused by a lack of objects with an infrared
excess. This is expected when one interprets the infrared emission as
arising from star formation, which in turn generates radio emission
according to the radio-infrared relation. \cite{Appleton2004} excluded
AGN based on spectroscopic observations and thus their sample is not
contaminated by AGN, and they do not see the tail towards low values
of $q_{24}$.

We note that the distribution of $q_{24}$ found by \cite{Norris2006a}
has a different shape than ours. It is rather constant between
$q_{24}=-0.5$ and $q_{24}=1.5$, and indicates a double-peaked
distribution. However, the differences in sensitivity make it
impossible to construct similar samples from the CDFS data presented
by \cite{Norris2006a} and the data presented here. We therefore
postpone a detailed analysis of the distribution of $q_{24}$ to the
time when the ATLAS survey is complete.

\section{Conclusions}
\label{sec:conc}

We have presented the first data from the ATLAS observations of the
ELAIS-S1 region, and a list of 1276 radio sources extracted from the
image. Radio sources have been matched to infrared SWIRE sources and
been classified as AGN if the morphology, radio-to-infrared ratio, or
the literature indicated so. We discover another 31 Infrared-Faint
Radio Sources, bringing the total number of these objects found with
the ATLAS survey to 55, and find no significant difference between the
distribution of source flux densities between the ELAIS-S1 and the CDFS
at $S_{\rm 20\um}>0.3\,{\rm mJy}$. We find a distribution of
$q_{24}={\rm log}(S_{\rm 20\um}/S_{\rm 20cm})$ which is in broad
agreement with the distribution found by
\cite{Appleton2004}. No further interpretation of our data is
presented, partly because other essential information such as
redshifts is not yet available and partly because the observations are
not yet complete.

\acknowledgments{The Australia Telescope Compact Array is operated by the CSIRO
Australia Telescope National Facility. IRS acknowledges support from
the Royal Society. This research has made extensive use of the
NASA/IPAC Extragalactic Database (NED) which is operated by the Jet
Propulsion Laboratory, California Institute of Technology, under
contract with the National Aeronautics and Space Administration.}

%\bibliography{refs}

\clearpage

\begin{figure}
\includegraphics[width=\linewidth]{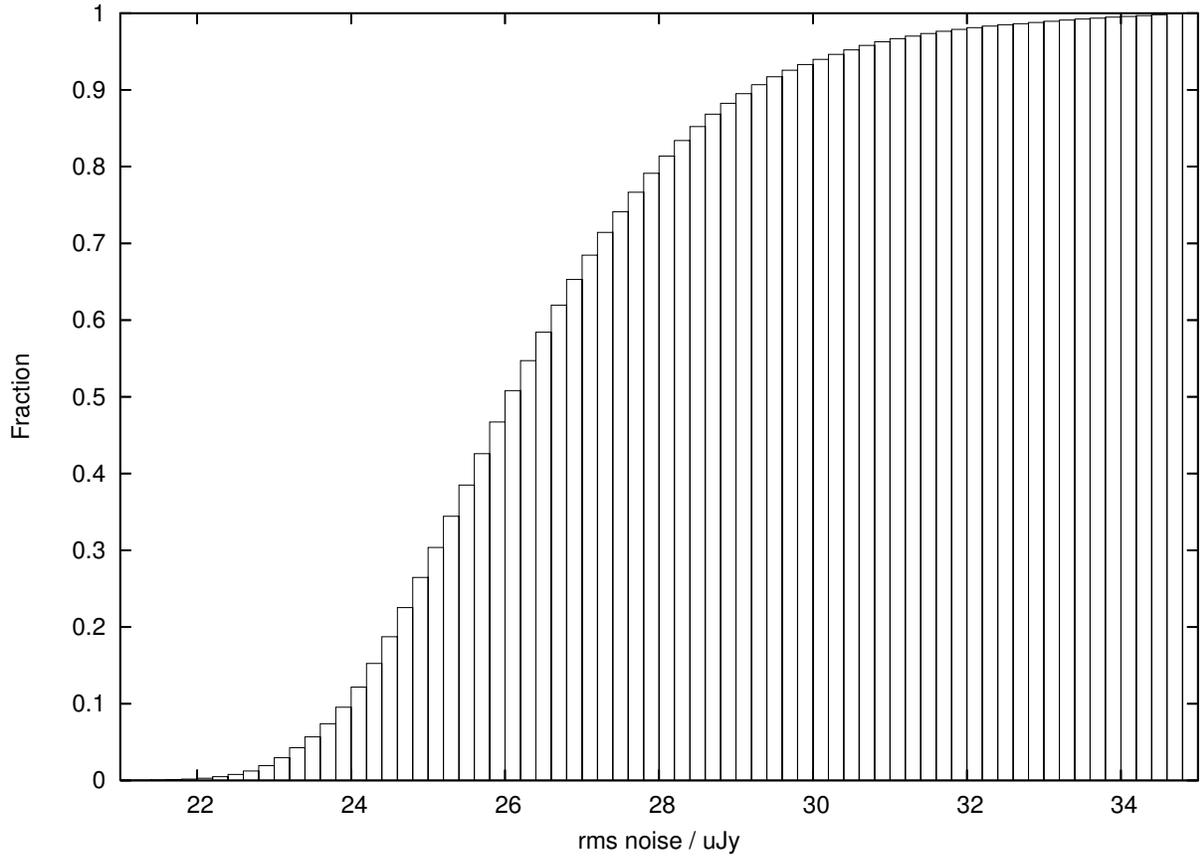}
\caption{Cumulative histogram of the pixel values of the rms map in the
central 1\,deg$^2$ of the observed area.}
\label{fig:cumhist}
\end{figure}

\clearpage

\begin{table}
\scriptsize
\begin{tabular}{lll}
\hline
\hline
Date    &   Config    & Int. time  \\
        &             & hours      \\
(1)     &  (2)        & (3)        \\
\hline
09,10,11 Jan 04   			& 6A      & 8.91, 8.77, 6.99  \\ 
30 Jan 04, 01 Feb 04 			& 6B      & 9.11, 9.47  \\
19, 27 Dec 04; 01, 02, 03 Jan 05 	& 1.5D    & 3.82, 9.09, 9.89, 8.41, 8.97  \\
09, 10, 11, 20, 21, 22 Jan 05 		& 750B    & 9.69, 9.51, 10.59, 4.15, 8.74, 7.29  \\
25 Mar 05; 08, 11 Apr 05		& 6A      & 8.9, 9.23, 9.02	  \\
24, 26, 30 Apr 05; 01 May 05		& 750A    & 8.16, 8.9, 8.74, 8.53  \\
08, 09 Jun 05 				& EW367   & 9.28, 9.05  \\
19, 24 Jun 05 				& 6B      & 9.17, 9.3  \\
\hline
\end{tabular}
\caption{Observing dates, array configurations and net integration
times on ELAIS-S1 pointings.}
\label{tab:obs}
\end{table}

\begin{table}
\scriptsize
\begin{tabular}{lll}
\hline
\hline
Source / Pointing    &   RA        & Dec        \\
                     &             &            \\
(1)                  &  (2)        & (3)        \\
\hline
1934-638             & 19:39:25.02 & -63:42:45.62 \\
0022-423             & 0:24:42.99  &  -42:02:03.95 \\
1                    & 0:32:03.55  &  -43:44:51.24 \\
2                    & 0:31:10.95  &  -43:27:59.64 \\
3                    & 0:32:05.04  &  -43:11:18.84 \\
4                    & 0:33:51.29  &  -43:11:24.96 \\
5                    & 0:32:57.67  &  -43:28:09.00 \\
6                    & 0:33:50.79  &  -43:44:57.36 \\
7                    & 0:35:38.02  &  -43:44:57.36 \\
8                    & 0:34:44.40  &  -43:28:11.88 \\
9                    & 0:35:37.51  &  -43:11:24.96 \\                        
10                   & 0:37:23.76  &  -43:11:18.84 \\
11                   & 0:36:31.13  &  -43:28:09.00 \\
12                   & 0:37:25.25  &  -43:44:51.24 \\
13                   & 0:36:31.13  &  -44:01:42.84 \\
14                   & 0:37:25.25  &  -44:18:34.44 \\
15                   & 0:35:38.02  &  -44:18:34.44 \\
16                   & 0:34:44.40  &  -44:01:42.84 \\
17                   & 0:32:57.67  &  -44:01:42.84 \\
18                   & 0:33:50.79  &  -44:18:34.44 \\
19                   & 0:32:03.55  &  -44:18:34.44 \\
20                   & 0:31:10.95  &  -44:01:42.84 \\
\hline
\end{tabular}
\caption{Coordinates of the calibrators and the pointings depicted in
Figure~\ref{fig:overview}.}
\label{tab:coords}
\end{table}

\begin{table}
\begin{tabular}{lrrr}
\hline
\hline
Separation & N & m & \%\\
(1)     &  (2)        & (3)        \\
\hline
$<1''   $  & 656   & 16    & 2.4\\
$1''-2''$  & 350   & 20    & 4.2\\
$2''-3''$  & 86    & 9     & 10.5\\
$>3''   $  & 45    & -     & -\\
\hline
\end{tabular}
\caption{Summary of the upper limits on the number of false
cross-identifications. Column 1 gives the separation, column 2 the
number of sources within this range, column 3 the number of radio
sources likely to be wrongly cross-identified with infrared sources,
and column 4 gives this number as a percentage.}
\label{tab:fxids}
\end{table}

\begin{deluxetable}{llllrrrrrrrrrr}
\tabletypesize{\scriptsize}
\tablewidth{0pt}
\rotate
\tablecaption{Radio component data\label{tab:radio1}}
\tablehead{
                &\colhead{Name} 	      & \colhead{RA}  	& \colhead{Dec}    & \colhead{RA err} & \colhead{Dec err} & \colhead{Peak} & \colhead{err} & \colhead{Int} & \colhead{err}    & \colhead{rms}     & \colhead{$B_{\rm maj}$} & \colhead{$B_{\rm min}$} & \colhead{PA}\\
                &               	      &               	&                  & \colhead{arcsec} & \colhead{arcsec}  & \colhead{mJy}  & \colhead{mJy} & \colhead{mJy} & \colhead{mJy}    & \colhead{\uJy\ } & \colhead{arcsec}        & \colhead{arcsec}        & \colhead{$^\circ$}\\
\colhead{(1)}   & \colhead{(2)}               & \colhead{(3)} 	& \colhead{(4)}    & \colhead{(5)}   & \colhead{(6)}      & \colhead{(7)}    & \colhead{(8)}      & \colhead{(9)}    & \colhead{(10)}   &  \colhead{(11)}   & \colhead{(12)}          & \colhead{(13)}          & \colhead{(14)}}
\startdata
C75 		& ATELAIS J003419.30-442647.2 & 00:34:19.308302 & -44:26:47.213520 & 0.33 & 0.23 & 0.25 & 0.03 & 0.25 & 0.01 & 31 & 10.26 & 7.17  & 0   \\
C76 		& ATELAIS J003247.08-442628.8 & 00:32:47.088391 & -44:26:28.830840 & 0.20 & 0.16 & 0.39 & 0.04 & 1.02 & 0.03 & 42 & 16.73 & 11.48 & 151 \\
C77 		& ATELAIS J003138.76-442620.6 & 00:31:38.765112 & -44:26:20.670360 & 0.16 & 0.12 & 0.35 & 0.04 & 0.35 & 0.01 & 39 & 10.26 & 7.17  & 0   \\
C78 		& ATELAIS J003152.54-442620.6 & 00:31:52.548125 & -44:26:20.666040 & 0.14 & 0.10 & 0.40 & 0.04 & 0.40 & 0.01 & 40 & 10.26 & 7.17  & 0   \\
C79 		& ATELAIS J003248.60-442625.7 & 00:32:48.606058 & -44:26:25.750680 & 0.38 & 0.26 & 0.31 & 0.04 & 0.31 & 0.02 & 42 & 10.26 & 7.17  & 0   \\
C80 		& ATELAIS J003659.30-442622.2 & 00:36:59.305858 & -44:26:22.295400 & 0.26 & 0.29 & 0.17 & 0.04 & 0.37 & 0.02 & 40 & 14.09 & 10.97 & 126 \\
C81 		& ATELAIS J003320.05-442617.8 & 00:33:20.053469 & -44:26:17.850480 & 0.09 & 0.07 & 0.42 & 0.04 & 0.47 & 0.01 & 39 & 10.79 & 7.63  & 21  \\
C82 		& ATELAIS J003832.11-442540.6 & 00:38:32.113102 & -44:25:40.639080 & 0.12 & 0.08 & 1.26 & 0.06 & 1.26 & 0.02 & 62 & 10.26 & 7.17  & 0   \\
C83 		& ATELAIS J003052.17-442537.3 & 00:30:52.170276 & -44:25:37.398360 & 0.28 & 0.20 & 0.29 & 0.05 & 0.48 & 0.02 & 55 & 15.14 & 7.89  & 30  \\
C84 		& ATELAIS J003253.48-442543.5 & 00:32:53.487876 & -44:25:43.583880 & 0.09 & 0.06 & 0.39 & 0.03 & 0.56 & 0.01 & 36 & 12.66 & 8.35  & 2   \\
C85 		& ATELAIS J003836.66-442513.5 & 00:38:36.662047 & -44:25:13.595520 & 0.06 & 0.05 & 1.66 & 0.06 & 2.57 & 0.03 & 67 & 12.04 & 9.46  & 165 \\
C86 		& ATELAIS J003602.72-442539.8 & 00:36:02.721341 & -44:25:39.837720 & 0.06 & 0.05 & 1.31 & 0.04 & 1.50 & 0.02 & 40 & 10.56 & 7.99  & 177 \\
C87 		& ATELAIS J003757.04-442516.6 & 00:37:57.045794 & -44:25:16.619160 & 0.41 & 0.29 & 0.28 & 0.04 & 0.28 & 0.02 & 44 & 10.26 & 7.17  & 0   \\
C88 		& ATELAIS J003543.38-442534.9 & 00:35:43.389367 & -44:25:34.921200 & 0.24 & 0.15 & 0.20 & 0.04 & 0.20 & 0.01 & 38 & 10.26 & 7.17  & 0   \\
C89 		& ATELAIS J003215.03-442521.8 & 00:32:15.038647 & -44:25:21.858960 & 0.24 & 0.22 & 0.21 & 0.03 & 0.29 & 0.02 & 37 & 10.97 & 9.24  & 152 \\
\enddata
\tablecomments{A section of the table with component data. Table~\ref{tab:radio1} is
published in its entirety in the electronic edition of the
Astronomical Journal.  A portion is shown here for guidance regarding
its form and content.  {\it Column 1:} a component number we use in
this paper. In some cases, sources were split up into sub-components,
resulting in component numbers such as ``C5'' and ``C5.1''. However,
this is no anticipation of the grouping of components to sources,
which was carried out independently; {\it column 2:} designation for
this component. In the case of single-component sources, this is
identical to the source name used in table 5. This is the formal IAU
designation and should be used in the literature when referring to
this component; {\it columns 3/4:} right ascension and declination
(J2000.0); {\it columns 5/6:} uncertainties in Right Ascension and
Declination. These include the formal uncertainties derived from the
Gaussian fit together with a potential systematic error in the
position of the calibrator source of 0.1 arcsec; {\it columns 7/8:}
peak flux density at 20\,cm (in mJy) of the fitted Gaussian component,
and the associated error as described in the text; {\it columns 9/10:}
integrated flux density at 20\,cm (in mJy) of the fitted Gaussian
component, and the associated error; {\it column 11:} the value (in
\uJy) of the rms map generated by SExtractor at the position of the
component; {\it columns 12/13/14:} the FWHM of the major and minor
axes of the Gaussian in arcsec, and its position angle in degrees.}
\end{deluxetable}

\begin{deluxetable}{lrrrrc}
\tabletypesize{\scriptsize}
\tablewidth{0pt}
\tablenum{\ref{tab:radio1}}
\rotate
\tablecaption{Radio component data (continued)}
\tablehead{
               & \colhead{Dec. Peak} & \colhead{Dec. $B_{\rm maj}$} & \colhead{Dec. $B_{\rm min}$} &\colhead{Dec. PA}   & \colhead{sidelobe?}\\
               & \colhead{mJy}       & \colhead{arcsec}             & \colhead{arcsec}             & \colhead{$^\circ$} \\
\colhead{(1)}  & \colhead{(15)}      & \colhead{(16)}               & \colhead{(17)}               & \colhead{(18)}     & \colhead{(19)}}
\startdata
C75 		\\
C76 		& 0.68 & 13.82 & 7.99 & 141 &   \\
C77 		\\
C78 		\\
C79 		\\
C80 		& 0.44 & 11.61 & 5.25 & 112 & * \\
C81 		\\
C82 		\\
C83 		\\
C84 		& 1.3 & 7.43 & 4.27 & 4 &  	  \\ 
C85 		& 5.21 & 7.27 & 4.99 & 129 &    \\
C86 		& 13.12 & 3.64 & 2.30 & 108 &   \\
C87 		\\
C88 		\\
C89 		& 2.15 & 6.85 & 1.45 & 110 &    \\
\enddata
\tablecomments{Radio component data (continued). {\it Column 15:} the
deconvolved peak flux density of the component in mJy. If the
undeconvolved fitted major or minor axis size was within one formal
standard error of the restoring beam size, no value is given; {\it
columns 16/17/18:} the deconvolved FWHM major and minor axes of the
Gaussian in arcsec, and its position angle in degrees. If the
undeconvolved fitted major or minor axis size was within one formal
standard error of the restoring beam size, no value is given; {\it
column 19:} an asterisk in this column indicates that this component
was deemed to be a sidelobe.}
\end{deluxetable}

\begin{deluxetable}{llp{7mm}lllrr}
\tabletypesize{\scriptsize}
\tablewidth{0pt}
\rotate
\tablecaption{Radio source data \label{tab:source1}}
\tablehead{
                   & \colhead{Name}  & \colhead{Comp.} & \colhead{RA}  & \colhead{Dec} &\colhead{SWIRE source} & \colhead{$S_{\rm 20cm}$}& \colhead{$\Delta S_{\rm 20cm}$}\\
                   &                 &                 &               &               &                       & \colhead{mJy}           & \colhead{mJy}                  \\
  \colhead{(1)}    & \colhead{(2)}   & \colhead{(3)}   & \colhead{(4)} & \colhead{(5)} & \colhead{(6)}         & \colhead{(7)}           & \colhead{(8)}                  }
\startdata
S693	& ATELAIS J003320.72-434030.1	& C693		& 00:33:20.72	& -43:40:30.11	& SWIRE4\_J003320.74-434030.1 	& 0.38	& 0.05	\\
S694	& ATELAIS J003020.95-433942.8	& C694		& 00:30:20.95	& -43:39:42.89	& SWIRE4\_J003020.97-433942.7 	& 49.58	& 2.48	\\
S695	& ATELAIS J003414.72-434030.7	& C695		& 00:34:14.72	& -43:40:30.74	& 	 			& 0.15	& 0.03	\\
S696	& ATELAIS J003402.27-434008.6	& C696		& 00:34:02.27	& -43:40:08.60	& SWIRE4\_J003402.20-434014.8 	& 0.49	& 0.04	\\
S697	& ATELAIS J003841.55-433925.0	& C697, C697.1	& 00:38:41.55	& -43:39:25.06	& SWIRE4\_J003841.54-433925.0 	& 13.32	& 0.67	\\
S698	& ATELAIS J003412.39-434005.8	& C698		& 00:34:12.39	& -43:40:05.84	& SWIRE4\_J003412.32-434005.2 	& 0.16	& 0.03	\\
S699	& ATELAIS J003513.86-433959.1	& C699		& 00:35:13.86	& -43:39:59.19	& SWIRE4\_J003513.86-433959.0 	& 0.33	& 0.04	\\
S700	& ATELAIS J003703.48-433935.5	& C700		& 00:37:03.48	& -43:39:35.56	& SWIRE4\_J003703.00-433935.3 	& 0.41	& 0.06	\\
S701	& ATELAIS J003141.08-433917.2	& C701		& 00:31:41.08	& -43:39:17.22	& SWIRE4\_J003141.18-433916.8 	& 0.50	& 0.07	\\
S702	& ATELAIS J003038.12-433903.8	& C702, C710	& 00:30:38.12	& -43:39:03.89	& SWIRE4\_J003038.11-433903.8 	& 1.48	& 0.10	\\  	
S703	& ATELAIS J003616.52-433917.5	& C703		& 00:36:16.52	& -43:39:17.55	& SWIRE4\_J003616.54-433918.3 	& 13.83	& 0.69	\\
S704	& ATELAIS J003544.33-433930.2	& C704		& 00:35:44.33	& -43:39:30.25	& SWIRE4\_J003544.38-433930.4 	& 0.22	& 0.04	\\
S705	& ATELAIS J003517.65-433931.9	& C705		& 00:35:17.65	& -43:39:31.97	& SWIRE4\_J003517.66-433931.0 	& 0.18	& 0.03	\\
S706	& ATELAIS J003815.05-433906.5	& C706		& 00:38:15.05	& -43:39:06.53	& SWIRE4\_J003814.95-433907.5 	& 0.34	& 0.07	\\
S707	& ATELAIS J003828.03-433847.2	& C707, C713	& 00:38:28.03	& -43:38:47.26	& SWIRE4\_J003828.02-433847.2 	& 6.00	& 2.44	\\

\enddata
\tablecomments{A section of the table with radio source data. Table~\ref{tab:source1} is
published in its entirety in the electronic edition of the
Astronomical Journal.  A portion is shown here for guidance regarding
its form and content.  {\it Column 1:} source number we use in this
paper; {\it column 2:} designation for this source. In the case of
single-component sources, this is identical to the component name used
in Table 3. This is the formal IAU designation and should be used in
the literature when referring to this source; {\it column 3:}
components which are deemed to belong to this source; {\it columns
4/5:} right ascension and declination (J2000.0). In the case of
single-component sources, this is the radio position of the
component. In the case of multi-component sources with good infrared
identification, the SWIRE position is used. In the case of
multi-component sources without infrared identification, the
coordinates are a flux-weighted mean of the components' coordinates;
{\it column 6:} name of the SWIRE source; {\it columns 7/8:}
integrated radio flux density of the source in mJy and the associated
error. In the case of extended or multiple-component sources, the flux
density has been integrated over the source region, rather than taking
the sum of its constituent components}
\end{deluxetable}

\begin{deluxetable}{lrrrrrrrrccrcp{6cm}}
\tabletypesize{\scriptsize}
\tablewidth{0pt}
\tablenum{\ref{tab:source1}}
\rotate
\tablecaption{Radio source data (continued)}
\tablehead{
       	       & \colhead{$S_{\rm 3.6\um}$}  & \colhead{$S_{\rm 4.5\um}$}    & \colhead{$S_{\rm 5.8\um}$}  & \colhead{$S_{\rm 8.0\um}$}  & \colhead{$S_{\rm 24\um}$}& \colhead{B}   & \colhead{V}   & \colhead{R}   & \colhead{AGN} & \colhead{M}    & \colhead{z}    & \colhead{ref}  & \colhead{comment}\\
       	       & \colhead{\uJy}              & \colhead{\uJy}                & \colhead{\uJy}              & \colhead{\uJy}              & \colhead{\uJy}           & \colhead{mag} & \colhead{mag} & \colhead{mag}\\
\colhead{(1)}  & \colhead{(9)}               & \colhead{(10)}                & \colhead{(11)}              & \colhead{(12)}              & \colhead{(13)}           & \colhead{(14)}& \colhead{(15)}& \colhead{(16)}& \colhead{(17)}& \colhead{(18)} & \colhead{(19)} & \colhead{(20)} & \colhead{(21)} }
\startdata
S693    & 253.94        & 275.32        & 281.19        & 561.27        & 2892.44         & 21.16   & 19.95   & 19.08   &        &          &         &         &                                                                                                                       \\
S694    & 435.13        & 312.91        & 158.69        & 86.96         &                 &         &         &         &        &          &         &         &                                                                                                                       \\                                        
S695    &               &               &               &               &                 &         &         &         &        &          &         &         & unclear XID                                                                                                           \\
S696    & 7.22          & 13.64         &               &               &                 &         &         &         &        &          &         &         & IFRS                                                                                                                  \\                                        
S697    & 155.47        & 103.93        & 94.40         & 49.86         & 233.13          & 22.39   & 21.53   & 20.51   & mf     &  x/-     &         &         & clearly a radio double, M-test fails due flux ratio of constituents                                                   \\  
S698    & 20.21         & 27.24         &               & 186.36        &                 & 24.68   & 24.78   & 24.04   &        &          &         &         &                                                                                                                       \\
S699    & 2038.05       & 1476.46       & 844.10        & 639.00        & 156.85          & 18.02   & 16.65   & 15.99   & f      &          & 0.11    & 6dF     &                                                                                                                       \\                                                        
S700    & 39.99         & 26.94         &               & 247.30        &                 & 22.69   & 22.04   & 21.23   & f      &          &         &         &                                                                                                                       \\
S701    & 145.32        & 83.23         & 58.80         &               &                 &         &         &         &        &          &         &         &                                                                                               			\\
S702    & 64.81         & 57.32         &               &               &                 &         &         &         & m      &          &         &         & extended, low surface brightness object, bridge of emission towards C710, which has no XID, hence core-jet morphology \\
S703    & 63.27         & 68.60         & 63.05         &               &                 & 25.14   &         &         &        &          &         &         & confused XID                                                                                                          \\                                
S704    & 14.14         & 15.81         &               &               &                 & 24.22   & 24.11   & 23.73   &        &          &         &         &                                                                                                                       \\
S705    & 45.86         & 48.11         & 54.40         & 655.28        &                 & 24.03   & 23.90   &         &        &          &         &         & confused XID                                                                                                          \\
S706    & 33.36         & 38.41         & 64.01         & 757.86        &                 & 24.85   & 24.39   & 24.07   &        &          &         &         & confused XID                                                                                                          \\
S707    & 6277.45       & 4263.60       & 9520.63       & 40203.53      & 27526.39        & 15.81   & 15.22   & 14.71   &        &          &         &         & C713 probably associated                                                                                              \\

\enddata
\tablecomments{Radio source data (continued). {\it Column 9-13:} flux density
of the infrared counterpart in the four IRAC bands at
3.6\,\um--8.0\,\um\ and in the MIPS band at 24\,\um, in
\uJy. Aperture-corrected flux densities have been used unless the
source was clearly extended, in which case the flux in a Kron aperture
has been used; {\it column 14-16:} optical magnitude of the
counterpart; {\it column 17:} flag indicating whether a source has
been classified as AGN or not, and based on what criteria. An ``f''
indicates AGN classification based on the far-infrared-radio relation,
an ``m'' based on morphology, and an ``l'' based on classification
taken from the literature; {\it column 18:} result of the test
developed by
\cite{Magliocchetti1998} as described in the text, performed for
double radio sources. A ``-'' indicates failure, a ``x'' success of
the two parts of the test (separation and flux density ratio of the
constituents); {\it column 19:} source redshift; {\it column 20:}
reference for the redshift. The codes indicate the following
publications: 2df - \cite{Colless2001}, 6dF -
\cite{Jones2004}, A01 - \cite{Alexander2001}, L04 - \cite{LaFranca2004}, 
P06 - \cite{Puccetti2006}, S01 - \cite{Serjeant2001}, S96 - \cite{Shectman1996},
W03 - \cite{Wegner2003}; {\it column 21:} comment}
\end{deluxetable}

\begin{deluxetable}{lll}
\tabletypesize{\scriptsize}
\tablewidth{0pt}
\tablenum{\ref{tab:source1}}
\rotate
\tablecaption{Radio source data (continued)}
\tablehead{
       	       & \colhead{G99 name} & \colhead{other name}\\
\colhead{(1)}  & \colhead{(22)}     & \colhead{(23)} }
\startdata
S693    & ELAIS20R\_J003021-433943 &                          \\                     
S694    &                          &                          \\                     
S695    & ELAIS20R\_J003402-434011 &                          \\                     
S696    & ELAIS20R\_J003842-433924 &                          \\                     
S697    &                          &                          \\                     
S698    &                          & 2MASX J00351384-4339588  \\                     
S699    &                          &                          \\                     
S700    &                          &                          \\                     
S701    &                          &                          \\                     
S702    & ELAIS20R\_J003617-433918 &                          \\                     
S703    &                          &                          \\                     
S704    &                          &                          \\                     
S705    &                          &                          \\                     
S706    & ELAIS20R\_J003828-433849 & ESO 242- G 021           \\                     
S707    & 			   &                          \\ 
\enddata
\tablecomments{Radio source data (continued). {\it Column 22:} The designation by \cite{Gruppioni1999}; {\it column 23:} Other names obtained from NED}
\end{deluxetable}

\begin{table}[htpb]
\scriptsize
\begin{tabular}{lllrrrp{5cm}}
\hline
\hline
Source 	& SUMSS RA        & SUMMS Dec       & $S$     &  Sep.  	& $\alpha$ & Comment \\
       	&                 &                 & mJy     & arcsec 	&          & \\
(1)    	& (2)             & (3)             & (4)     & (5)    	& (6)      & (7)\\
\hline
S207    & 00:30:48.60     & -44:14:33.10    & 34.6    & 14.40   & -1.06   & S207, S207.2 and S212 all blend together in this source \\
S220    & 00:37:09.57     & -44:14:08.40    & 11.6    & 2.18    & -1.48   &							    \\
S258    & 00:32:04.54     & -44:11:32.70    & 59.0    & 2.36    & -1.00   &							    \\
S272    & 00:36:44.04     & -44:10:54.90    & 13.3    & 2.53    & -0.96   & blends with S278					    \\
S293    & 00:29:25.72     & -44:08:24.80    & 12.4    & 2.08    & -1.11   & blends with S304					    \\
S296    & 00:36:50.07     & -44:08:59.70    & 14.7    & 3.00    & -0.80   &							    \\
S311    & 00:37:20.40     & -44:07:31.20    & 74.3    & 4.34    & -0.78   &							    \\
S313    & 00:31:10.76     & -44:07:41.70    & 14.4    & 1.13    & -1.01   &							    \\
S325    & 00:34:52.73     & -44:07:26.30    & 10.8    & 1.31    & -1.13   &							    \\
S347    & 00:35:38.62     & -44:06:03.60    & 15.9    & 15.98   & -0.20   &							    \\
S355    & 00:30:19.00     & -44:04:33.40    & 14.9    & 5.67    & -0.70   &							    \\
S360    & 00:34:58.74     & -44:04:59.30    & 26.2    & 1.95    & -0.53   &							    \\
S371    & 00:38:54.63     & -44:03:29.20    & 12.1    & 0.99    & -0.37   &							    \\
S381    & 00:39:40.19     & -44:02:10.20    & 23.1    & 2.20    &  0.19   & 							    \\
S390.1  & 00:37:19.70     & -44:01:49.80    & 11.0    & 3.64    & -3.13   & blends with 390                                         \\
\hline
\end{tabular}
\caption{A section of the table with SUMSS counterparts to 1.4\,GHz
radio sources. Table~\ref{tab:sumss} is published in its entirety in
the electronic edition of the Astronomical Journal.  A portion is
shown here for guidance regarding its form and content. {\it Column
1:} The source names we use in this paper; {\it Columns 2/3:} SUMMS
right ascension and declination; {\it Column 4:} SUMMS flux density in
mJy; {\it Column 5:} separation of the SUMMS source to the source
coordinates in Table~\ref{tab:source1}; {\it Column 6:} the spectral
index; {\it Column 7:}  comment}
\label{tab:sumss}
\end{table}

\end{document}